\documentclass[12pt]{article}
\usepackage{bm}
\usepackage{a4wide,graphicx}
\usepackage{cite}
\begin{document}


\title{Strange mesons in nuclear matter at finite temperature}

\author{L. Tol\'os $^{1}$, D. Cabrera$^2$ and A. Ramos$^3$\\
$^1$ FIAS.
Goethe-Universit\"at Frankfurt am Main,\\
Ruth-Moufang-Str. 1, 60438 Frankfurt am Main, Germany\\
$^2$Departamento de F\'{\i}sica Te\'orica II, Universidad Complutense,\\
28040 Madrid, Spain\\
$^3$ Departament d'Estructura i Constituents de la Mat\`eria\\
Universitat de Barcelona,
Diagonal 647, 08028 Barcelona, Spain}

\date{\today}

\maketitle
\begin{abstract}

We study the properties of $K$ and $\bar K$ mesons in nuclear matter at finite
temperature from a chiral unitary approach in coupled channels which
incorporates the $s$- and $p$-waves of the kaon-nucleon interaction. The
in-medium solution accounts for Pauli blocking effects, mean-field binding on
all the baryons involved, and $\pi$ and kaon self-energies. We calculate $K$
and $\bar K$ (off-shell) spectral functions and single particle properties. The
$\bar K$ effective mass gets lowered by about $-50$~MeV in cold nuclear matter
at saturation density and by half this reduction at $T=100$~MeV. The $p$-wave
contribution to the ${\bar K}$ optical potential, due to $\Lambda$, $\Sigma$
and $\Sigma^*$ excitations, becomes significant for momenta larger than 200
MeV/c and reduces the attraction felt by the $\bar K$ in the nuclear medium.
The $\bar K$ spectral function spreads over a wide range of energies,
reflecting the melting of the $\Lambda (1405)$ resonance and the contribution
of $YN^{-1}$ components at finite temperature.  In the $KN$ sector, we find
that  the low-density theorem is a good approximation for the $K$ self-energy
close to saturation density due to the absence of resonance-hole excitations.
The $K$ potential shows a moderate repulsive behavior, whereas the
quasi-particle peak is considerably broadened with increasing density and
temperature. 
We discuss the implications for the decay of the $\phi$ meson at SIS/GSI energies
as well as in the future FAIR/GSI project.

\end{abstract}
\vskip 0.5 cm



\section{Introduction}
\label{sec:intro}

The properties of hadrons and, in particular, of mesons with strangeness in
dense matter have been a matter of intense investigation over the last years, in
connection to the study of exotic atoms \cite{Friedman:2007zz} and the analysis
of heavy-ion collisions (HIC's) \cite{Rapp:1999ej}.

At zero temperature, the study of the $\bar K$ interaction in
nuclei has revealed some interesting characteristics. First, 
the presence below the
$\bar K N$ threshold of the $\Lambda(1405)$ resonance gives rise to
the failure of the $T \rho$ approximation for the $\bar K$
self-energy. Whereas the $\bar K N$ interaction is repulsive at
threshold, the phenomenology of kaonic atoms requires an attractive
potential. The consideration of Pauli blocking on the intermediate
$\bar K N$ states \cite{Koch,schaffner} was found to shift the
excitation energy of the $\Lambda (1405)$ to higher energies,
hence changing the real part of the $\bar K N$ amplitude from
repulsive in free space to attractive in a nuclear medium already
at very low densities. Further steps were taken to account for
self-consistency in the evaluation of the $\bar K$ self-energy
\cite{Lutz} as well as for relevant medium effects on the
intermediate meson-baryon states \cite{Ramos:1999ku}, which
results in a moderate final size of the attractive potential as
well as in a sizable imaginary part associated to several in-medium decay
mechanisms \cite{Ramos:1999ku,TOL00,Tolos:2002ud,TOL06}. 
Several studies of the $\bar K$ potential based on phenomenology
of kaonic atoms have pointed towards a different class of deeply
attractive potentials \cite{gal}. Unfortunately, the present
experimental knowledge is unable to solve this controversy, as
both kinds of potentials fairly describe the data from kaonic
atoms \cite{gal,baca}.

A different direction in the study of this problem was given in \cite{akaishi},
where a highly attractive antikaon-nucleus potential was constructed leading to
the prediction of narrow strongly bound states in few body systems
\cite{akaishi,dote04,akaishi05}. This potential has been critically discussed
in \cite{toki} because of the omission of the direct coupling of the $\pi
\Sigma$ channel to itself, the assumption of the nominal $\Lambda(1405)$ as a
single bound $\bar{K}$ state, the lack of self-consistency in the calculations
and the seemingly too large nuclear densities obtained of around ten times
normal nuclear matter density at the center of the nucleus. Experiments devised
to the observation of deeply bound kaonic states, measuring particles emitted
after the absorption of $K^-$ in several nuclei, reported signals that could
actually be interpreted from conventional nuclear physics processes. The
experimental observations could be explained simply either by the two-body
absorption mechanism raised in \cite{toki}, without
\cite{kek1,kek2,finuda1,magas2} or with \cite{finuda2,magas1} final state
interactions, or coming from a three-body absorption process
\cite{kek3,finuda3,magas3}. Actually, recent improved few-body calculations using
realistic ${\bar K} N$ interactions and short-range correlations
\cite{shevchenko,shevchenko2,Ikeda:2007nz,dote_hyp06} predict 
few-nucleon kaonic states bound only by 50--80 MeV
and having large widths of the order on 100 MeV, thereby disclaiming the
findings of Refs.~\cite{akaishi,dote04,akaishi05}.

Relativistic heavy-ion experiments at beam energies below 2AGeV
\cite{Forster:2007qk,FOPI} is another experimental scenario that
has been testing the properties of strange mesons not only in a
dense but also in a hot medium. Some interesting conclusions have
been drawn comparing the different theoretical transport-model
predictions and the experimental outcome on production cross
sections, and energy and polar-angle distributions
\cite{Forster:2007qk}. For example, despite the significantly
different thresholds in binary $NN$ collisions, there is a clear
coupling between the $K^-$ and $K^+$ yields since the $K^-$ is
predominantly produced via strangeness exchange from hyperons
which, on the other hand, are created together with $K^+$ mesons.
Furthermore, the $K^-$ and $K^+$ mesons exhibit different
freeze-out conditions as the $K^-$ are continuously produced and
reabsorbed, leaving the reaction zone much later than the $K^+$
mesons. However, there is still not a consensus on the influence
of the kaon-nucleus potential on those observables
\cite{Cassing:2003vz}.

Compared to the $\bar K N$  interaction, the $KN$ system has
received comparatively less attention. Because of the lack of
resonant states in the $S=+1$ sector, the single-particle
potential of kaons has usually been calculated in a $T \rho$
approximation, with a repulsion of around 30 MeV for nuclear
matter density (the information on $T$ is taken from scattering
lengths and energy dependence is ignored). However, a recent
analysis of the $K N$ interaction in the J\"ulich model has
demonstrated  that the self-consistency induces  a significant
difference in the optical potential with respect to the
low-density approximation at saturation density, and the kaon
potential exhibits a non-trivial momentum dependence
\cite{Tolos:2005jg}.

A precise knowledge of kaon properties in a hot and dense medium
is also an essential ingredient to study the fate of the $\phi$
meson. Electromagnetic decays of vector mesons offer a unique
probe of high density regions in nuclear production experiments
and HIC's \cite{Rapp:1999ej}. The $\phi$ meson predominantly decays into $\bar K K$,
which are produced practically at rest in the center of mass
frame, each carrying approximately half of the mass of the vector
meson. Such a system is highly sensitive to the available phase
space, so that small changes in the kaon effective masses or the
opening of alternative baryon-related decay channels may have a
strong repercussion in the $\phi$-meson mass and decay width. The
analysis of the mass spectrum of the $\phi$ decay products in
dedicated experiments has drawn inconclusive results since the
long-lived vector meson mostly decays out of the hot/dense system
\cite{Akiba:1996ab,Adler:2004hv,Adamova:2005jr,Ishikawa:2004id,Muto:2006eg,:2007mga}.
In addition, despite the sizable modifications predicted in most
theoretical studies
\cite{Hatsuda:1991ez,Asakawa:1994tp,Zschocke:2002mn,Klingl:1997tm,Oset:2000eg,Cabrera:2002hc,Smith:1997xu,AlvarezRuso:2002ib},
the current experimental resolution typically
dominates the observed spectrum from $\phi$ decays.
Still, recent measurements of the
$\phi$ transparency ratio in the nuclear photoproduction reaction
by the LEPS Collaboration \cite{Ishikawa:2004id}
have shed some light on the problem and seem to
indicate an important renormalization of the absorptive part of
the $\phi$ nuclear potential, as it was suggested in the
theoretical analysis of Ref.~\cite{Cabrera:2003wb}. The
unprecedented precision achieved by the CERN NA60 Collaboration in
the analysis of dimuon spectrum data from In-In collisions at
158~AGeV \cite{Arnaldi:2006jq}, as well as the advent of future studies 
of vector meson
spectral functions to be carried out at the HADES \cite{HADES} and
CBM \cite{CBM} experiments in the future FAIR facility, advise an
extension of our current theoretical knowledge of the $\phi$
spectral function to the $(\mu_B, T)$ plane, and hence, of the $K$
and $\bar K$ properties at finite temperature and baryon density.

In this work we evaluate the $K$ and $\bar K$ self-energy, spectral
function and nuclear optical potentials in a nuclear medium at
finite temperature. We follow the lines of
Refs.~\cite{Ramos:1999ku,TOL06,TOL07,Oset:1997it} and build the
$s$-wave $K N$ and $\bar K N$ $T$-matrix in a coupled channel
chiral unitary approach. Medium effects are incorporated by
modifying the intermediate meson-baryon states. We account for
Pauli blocking on intermediate nucleons, baryonic binding
potentials and meson self-energies for pions and kaons. The latter
demands a self-consistent solution of the $K$ and $\bar K$
self-energies as one sums the kaon nucleon scattering amplitude
over the occupied states of the system, whereas the $T$-matrix
itself incorporates the information of the kaon self-energies in
the intermediate meson-baryon Green's functions. The interaction
in $p$-wave is also accounted for in the form of $YN^{-1}$
excitations, which lead to a sizable energy dependence of the
$\bar K$ self-energy below the quasi-particle peak. Finite
temperature calculations have been done in the Imaginary Time
Formalism in order to keep the required analytical properties of
retarded Green's functions which, together with the use of
relativistic dispersion relations for baryons (and, of course, for
mesons), improves on some approximations typically used in former
works. The organization of the present article goes as follows: in
Sect.~\ref{sec:Form} we develop the formalism and ingredients on
which the calculation is based. Sect.~\ref{sec:Resul} is devoted
to the presentation of the results. The $\bar K$
and $K$ self-energies and spectral functions are discussed in
Sects.~\ref{ssec:Resul-Kbar-spectral} and
\ref{ssec:Resul-Kaon-spectral}, respectively. We devote Sect.~\ref{ssec:optical} 
to the discussion of momentum, density and
temperature dependence of $\bar K$ and $K$ nuclear optical
potentials. Finally, in Sect.~\ref{sec:Conclusion}  we draw our
conclusions as well as the implications of the in-medium properties of kaons at finite temperature
in transport calculations and $\phi$ meson phenomenology. We also 
give final remarks pertaining to the present and
future works.

\section{Kaon nucleon scattering in hot nuclear matter}
\label{sec:Form}

In this section we discuss the evaluation of the effective kaon nucleon
scattering amplitude in a dense nuclear medium, extending the unitarized chiral
model for $\bar K$ of Refs.~\cite{Ramos:1999ku,TOL06} to account for finite
temperature. This allows us to obtain the in-medium $K$ and ${\bar K}$
self-energy, spectral function and nuclear optical potential. We follow closely
the lines of Ref.~\cite{TOL07}, where a similar study was reported for
open-charm mesons in hot nuclear matter.

\subsection{$s$-wave kaon nucleon scattering and kaon self-energy}
\label{ssec:swave-kaon-self-energy}

The kaon nucleon interaction at low energies has been successfully described in
Chiral Perturbation Theory ($\chi$PT)
\cite{Gasser:1984gg,Meissner:1993ah,Bernard:1995dp,Pich:1995bw,Ecker:1994gg}.
The lowest order chiral Lagrangian which couples the octet of light pseudoscalar
mesons to the octet of $1/2^+$ baryons is given by
\begin{eqnarray}
   {\cal L}_1^{(B)} &=& \langle \bar{B} i \gamma^{\mu} \nabla_{\mu} B
    \rangle  - M \langle \bar{B} B\rangle  \nonumber \\
    &&  + \frac{1}{2} D \left\langle \bar{B} \gamma^{\mu} \gamma_5 \left\{
     u_{\mu}, B \right\} \right\rangle + \frac{1}{2} F \left\langle \bar{B}
     \gamma^{\mu} \gamma_5 \left[u_{\mu}, B\right] \right\rangle  \ ,
    \label{chiralLag}
\end{eqnarray}
where $B$ is the $SU(3)$ matrix for baryons, $M$ is the baryon mass, $u$
contains the $\Phi$ matrix of mesons and the symbol $\langle \, \rangle$ denotes
the flavour trace. The $SU(3)$ matrices appearing in Eq.~(\ref{chiralLag}) are
standard in notation and can be found, for instance, in \cite{Oset:1997it}. The
axial-vector coupling constants have been determined in \cite{Jido:2003cb} and
read $D=0.8$ and $F=0.46$.

Keeping at the level of two meson fields, the covariant derivative term in
Eq.~(\ref{chiralLag}) provides the following interaction Lagrangian in $s$-wave,
\begin{equation}
   {\cal L}_1^{(B)} = \left\langle \bar{B} i \gamma^{\mu} \frac{1}{4 f^2}
   [(\Phi\, \partial_{\mu} \Phi - \partial_{\mu} \Phi \Phi) B
   - B (\Phi\, \partial_{\mu} \Phi - \partial_{\mu} \Phi \Phi)]
   \right\rangle \ . \label{lowest}
\end{equation}
Expanding the baryon spinors and vertices in $M_B^{-1}$ the following expression
of the $s$-wave kaon nucleon tree level amplitude can be derived
\begin{eqnarray}
V_{i j}^s= - C_{i j} \, \frac{1}{4 f^2} \, (2 \, \sqrt{s}-M_{B_i}-M_{B_j})
\left( \frac{M_{B_i}+E_i}{2 \, M_{B_i}} \right)^{1/2} \, \left(
\frac{M_{B_j}+E_j}{2 \, M_{B_j}} \right)^{1/2} \ ,
\label{swa}
\end{eqnarray}
with $M_{B_i}$ and $E_i$ the mass and energy of the baryon in the $i$th channel,
respectively. The coefficients $C_{ij}$ form a symmetric matrix and are given in
\cite{Oset:1997it}. The meson decay constant $f$ in the $s$-wave amplitude is
taken as  $f=1.15 f_\pi$. This renormalized value provides a satisfactory
description of experimental low energy $\bar K N$ scattering observables (such
as cross sections and the properties of the $\Lambda (1405)$ resonance) using
only the interaction from the lowest order chiral Lagrangian plus unitarity in
coupled channels.
We have considered the following channels in the calculation of the scattering
amplitude: in the strangeness sector $S=-1$ we have $\bar K N$, $\pi \Sigma$,
$\eta \Lambda$ and $K \Xi$ for isospin $I=0$; $\bar K N$, $\pi \Lambda$, $\pi
\Sigma$, $\eta \Sigma$ and $K \Xi$ for $I=1$. For $S=1$, there is only one
single channel, $K  N$, for each isospin.

In Refs.~\cite{Oset:1997it,Ramos:1999ku,TOL06} unitarization of the tree level
amplitudes in coupled channels was implemented, which extends the applicability
of $\chi$PT to higher energies and in particular allows to account for
dynamically generated resonances. In particular, the $\Lambda(1405)$ shows up in
the unitarized $s$-wave ${\bar K} N$ amplitude. Following \cite{Oset:1997it},
the effective kaon-nucleon scattering amplitude is  obtained by solving the
Bethe-Salpeter equation in coupled channels (in matrix notation),
\begin{equation}
T =  V + \overline{V G T} \, ,
\end{equation}
where we use the $s$-wave tree level amplitudes as the potential (kernel) of the
equation, $V^s_{ij}$, and
\begin{equation}
\label{G_vacuum}
G_i (\sqrt{s}) = {\rm i} \,
\int \frac{d^4q}{(2\, \pi)^4} \,
\frac{M_i}{E_i(-\vec{q}\,)} \,
\frac{1}{\sqrt{s} - q_0 - E_i(-\vec{q}\,) + {\rm i} \varepsilon} \,
\frac{1}{q_0^2 - \vec{q}\,^2 - m_i^2 + {\rm i} \varepsilon}
\end{equation}
stands for the intermediate two-particle meson-baryon Green's
function of channel $i$ ($G$ is diagonal). In principle, both $V$
and $T$ enter off-shell in the momentum integration
($\overline{VGT}$ term) of the meson-baryon loop. However, as it
was shown in Refs.~\cite{Oset:1997it,TOL06}, the (divergent)
off-shell contributions of $V$ and $T$ in the $s$-wave interaction
can be reabsorbed in a renormalization of the bare coupling
constants and masses order by order. Therefore, both $V$ and $T$
can be factorized on-shell out of the meson-baryon loop, leaving
the four-momentum integration only in the two-particle
meson-baryon propagators. An alternative justification of solving
the Bethe-Salpether equation with on-shell amplitudes may be found
in the framework of the $N/D$ method, applied for meson-meson
interactions in Ref.~\cite{Oller:1998zr} and for meson-baryon
interactions in Ref.~\cite{Oller:2000fj}. We are thus left with a
set of linear algebraic equations with trivial solution,
\begin{equation}
T = [1 - V G ]^{-1} V \,\,\, .
\label{eq:BSalgeb}
\end{equation}
The meson baryon loop function, $G_i$, needs to be regularized. We apply a
cut-off in the three momentum of the intermediate particles, which provides a
simple and transparent regularization method for in-medium calculations,
cf.~\cite{Ramos:1999ku,TOL06}.

In order to obtain the effective $s$-wave $\bar K(K) N$ amplitude in hot and
dense matter, we incorporate in the loop functions the modifications
on the properties of the mesons and
baryons induced by temperature and density.

In the Imaginary Time Formalism (ITF), the baryon
propagator in a hot medium is given by:
\begin{equation}
{\cal G}_B(\omega_m,\vec{p};T) =
\frac{1}{{\rm i}\omega_m-E_B(\vec{p},T)}\ ,
\label{eq:nuc}
\end{equation}
where ${\rm i} \omega_m={\rm i} (2m+1)\pi T + \mu_B$
is the fermionic Matsubara frequency, with $\mu_B$ the baryon chemical
potential,
and $E_B$ is the baryon single particle energy, which, in the case
of nucleons and singly strangeness hyperons, will also contain the
medium binding effects obtained within a temperature dependent
Walecka-type $\sigma -\omega$ model (see Ref.~\cite{KAP-GALE}).
According to this model, the nucleon energy spectrum in mean-field
approximation is obtained from
\begin{eqnarray}
E_N(\vec{p},T)=\sqrt{\vec{p}\,^2+M_N^*(T)^2}+\Sigma^v \ ,
\end{eqnarray}
with the vector potential $\Sigma^v$ and the effective mass $M_N^*(T)$ given  by
\begin{eqnarray}
\Sigma^v&=&\left(\frac{g_v}{m_v}\right)^2 \rho    \nonumber \\
M_N^*(T)&=&M_N-\Sigma^s, ~~~~~~~~~{\rm with}~\Sigma^s=
\left(\frac{g_s}{m_s}\right)^2 \rho_s \ ,
\end{eqnarray}
where $m_s$ and $m_v$ are the meson masses ($m_s=440$~MeV, $m_v=782$~MeV), while
$g_s$ and $g_v$ are the scalar and vector density dependent
coupling constants.
These constants are obtained by reproducing the energy per particle of symmetric
nuclear matter at $T=0$ coming from a Dirac-Brueckner-Hartree-Fock calculation
(see  Table 10.9 of Ref.~\cite{Machleidt:1989tm}). The vector ($\rho$) and
scalar ($\rho_s$) densities are obtained by momentum integration, namely
\begin{eqnarray}
\rho_{(s)} = 4 \,\int  \frac{ d^3p}{(2\pi)^3} \,
n_N^{(s)}(\vec{p},T) \ , \label{eq:density}
\end{eqnarray}
of the corresponding vector [$n_N(\vec p, T)$] and scalar
[$n_N^s(\vec p, T)$] density distributions, which are defined in
terms of the nucleon Fermi-Dirac function as
\begin{equation}
n_N(\vec{p}, T)=\frac{1}{1+\exp{\left [(E_N(\vec{p},
T)-\mu_B)/T\right ]}} \label{eq:density-dist}
\end{equation}
and
\begin{equation}
n_N^s(\vec{p}, T)=\frac{M_N^*(T)n_N(\vec
p,T)}{\sqrt{\vec{p}\,^2+M_N^*(T)^2}}\ , \label{eq:density-dist-s}
\end{equation}
respectively. The
quantities $E_N(\vec{p}, T), M_N^*(T)$ and $\mu_B$ are obtained
simultaneously and self-consistently for given $\rho$ and $T$ and
for the corresponding values of $g_s$ and $g_v$.

The hyperon masses and energy spectra,
\begin{eqnarray}
E_{Y}(\vec{p},T)=\sqrt{\vec{p}\,^2+M_{Y}^*(T)^2}+\Sigma_{Y}^v \ ,
\end{eqnarray}
can be easily inferred from those for the
nucleon as
\begin{eqnarray}
\Sigma_{Y}^v&=&\frac{2}{3} \left(\frac{g_v}{m_v}\right)^2 \rho= \frac{2}{3}\Sigma^v  \nonumber \ ,\\
M_{Y}^*(T)&=&M_{Y}-\Sigma_{Y}^s=M_{Y}-\frac{2}{3}\left(\frac{g_s}{m_s}\right)^2 \rho_s \nonumber \\
&=& M_{Y}-\frac{2}{3}(M_N-M_N^*(T)) \ .
\end{eqnarray}
Here we have assumed that the $\sigma$ and $\omega$ fields only couple to the
$u$ and $d$ quarks, as in Refs.~\cite{Tsushima:2002cc,Tsushima:2003dd}, so the
scalar and vector coupling constants for hyperons and charmed baryons are
\begin{eqnarray}
g_v^{Y}=\frac{2}{3}g_v , \hspace{1cm} g_s^{Y}=\frac{2}{3}g_s .
\end{eqnarray}
In this way, the potential for hyperons follows the simple light quark counting
rule as compared with the nucleon potential: $V_{Y}=2/3 \, V_N$.
As reference, we quote in Table \ref{table:dmuB} the nucleon and hyperon
single particle properties for three densities ($0.25\rho_0$, $\rho_0$ and
$2\rho_0$), where $\rho_0=0.17$~fm$^{-3}$ is the normal nuclear matter saturation
density, and four temperatures ($T=0$, 50, 100 and 150~MeV).
The hyperon
attraction at $\rho=\rho_0$ and $T=0$ MeV is about  $-50$ MeV, the size of which
gets reduced as temperature increases turning even into repulsion, especially at
higher densities. This behavior results from the fact that the temperature
independent vector potential takes over the strongly temperature-dependent
scalar potential which decreases with temperature.
We note that the quark-meson coupling (QMC) calculations of
Refs.~\cite{Tsushima:2002cc,Tsushima:2003dd}, performed at $T=0$, obtained a
somewhat smaller scalar potential (about half the present one) for the $\Lambda$
and $\Sigma$  baryons due to the inclusion of non-linear terms associated to
quark dynamics. To the best of our
knowledge, no temperature effects have been studied within this framework.

\begin{table}[tb]
    \centering
    \caption{$\sigma-\omega$ model at finite temperature}
   \begin{tabular}{ccccccccccc}
      $\rho[{\rm fm}^{-3}]$ & T[MeV] & $\mu_B$[MeV] & $M_N^*$[MeV] & $\Sigma^v$[MeV] & $M_{\Lambda}^*$[MeV] & $\Sigma_{\Lambda}^v$[MeV] & $M_{\Sigma}^*$[MeV] & $\Sigma_{\Sigma}^v$[MeV]\\
\hline
              0.0425 &   0 & 920 & 781 & 121 &1010 &  81 &1088 &  81 \\
              0.0425 &  50 & 820 & 793 & 121 &1019 &  81 &1096 &  81 \\
              0.0425 & 100 & 618 & 805 & 121 &1026 &  81 &1104 &  81 \\
              0.0425 & 150 & 364 & 815 & 121 &1033 &  81 &1110 &  81 \\

              0.17 &   0 & 920 & 579 & 282 & 872 & 188 & 950 & 188 \\
              0.17 &  50 & 892 & 605 & 282 & 893 & 188 & 970 & 188 \\
              0.17 & 100 & 783 & 634 & 282 & 913 & 188 & 990 & 188 \\
              0.17 & 150 & 618 & 659 & 282 & 929 & 188 &1006 & 188 \\

              0.34 &   0 & 979 & 443 & 422 & 787 & 281 & 865 & 281 \\
              0.34 &  50 & 969 & 470 & 422 & 803 & 281 & 881 & 281 \\
              0.34 & 100 & 905 & 510 & 422 & 830 & 281 & 907 & 281 \\
              0.34 & 150 & 787 & 545 & 422 & 853 & 281 & 930 & 281 \\

    \end{tabular}
    \label{table:dmuB}
\end{table}

The meson propagator in a hot medium is given by
\begin{equation}
D_M(\omega_n,\vec{q};T) = \frac{1}{({\rm i} \omega_n)^2-\vec{q}\,^2 - m_M^2 -
\Pi_M(\omega_n,\vec{q};T)} \ ,
\label{eq:prop1}
\end{equation}
where ${\rm i} \omega_n = {\rm i} 2 n \pi T$
is the bosonic Matsubara frequence and
$\Pi_M (\omega_n, \vec{q}; T)$ is the meson self-energy.
Note that throughout this work we set the mesonic chemical potential to zero,
since we are dealing with an isospin symmetric nuclear medium with zero
strangeness.
We will consider in this work the dressing of pions and kaons.

An evaluation of the in-medium self-energy for pions at finite
temperature and baryonic density was given in the Appendix of
Ref.~\cite{Tolos:2002ud},  which generalized the zero temperature
evaluation of the pion self-energy from
Refs.~\cite{Oset:1989ey,Ramos:1994xy} by incorporating thermal
effects. We recall that the pion self-energy in nuclear matter at
$T=0$ is strongly dominated by the $p$-wave coupling to
particle-hole ($ph$) and $\Delta$-hole ($\Delta h$) components (a small,
repulsive $s$-wave contribution takes over at small momenta), as
well as to 2$p$-2$h$ excitations, which account for pion
absorption processes, and short range correlations. We come back
to the pion self-energy in Sec.~\ref{ssec:pion-self-energy}, where
we improve on some approximations in previous works.

In the case of the kaons, the self-energy receives contributions
of comparable size from both $s$- and $p$-wave interactions with
the baryons in the medium. We evaluate the $s$-wave self-energy
from the effective $\bar K (K)N$ scattering amplitude in the
medium, a procedure which, as will be shown explicitly in the
following, must be carried out self-consistently. The $p$-wave
part of the kaon self-energy will be discussed separately in the
next section.

The evaluation of the effective $\bar K (K)N$ $s$-wave scattering
amplitude in the hot medium proceeds by first obtaining the
meson-baryon two-particle propagator function (meson-baryon loop)
at finite temperature in the ITF, ${\cal G}_{MB}$. Given the
analytical structure of ${\cal G}_{MB}$ it is convenient to use
the spectral (Lehmann) representation for the meson propagator,
\begin{eqnarray}
\label{Lehmann}
D_M(\omega_n,\vec{q};T) &=& \int d\omega \,
\frac{S_M(\omega,\vec{q};T)}{{\rm i}\omega_n -
\omega}
\nonumber \\
&=&
\int_0^{\infty} d\omega \,
\frac{S_M(\omega,\vec{q};T)}{{\rm i}\omega_n - \omega}
-
\int_0^{\infty} d\omega \,
\frac{S_{\bar M}(\omega,\vec{q};T)}{{\rm i}\omega_n + \omega}
\,\,\,,
\end{eqnarray}
where 
$S_M$, $S_{\bar M}$ stand for the spectral functions of the
meson and its corresponding anti-particle.
The separation in the
second line of Eq.~(\ref{Lehmann}) reflects the retarded character
of the meson self-energy and propagator,  ${\rm Im} \, \Pi
(-q_0,\vec{q};T) = - {\rm Im} \, \Pi (q_0,\vec{q};T)$. Due to
strangeness conservation, $K$ and ${\bar K}$ experience markedly
different interactions in a nuclear medium \cite{Ramos:1999ku},
which justifies using a different notation ($S_M$, $S_{\bar M}$)
for the spectral functions in Eq.~(\ref{Lehmann})\footnote{In the
case of pions, for instance, in an isospin symmetric nuclear
medium, all the members of the isospin triplet ($\pi^\pm$,
$\pi^0$) acquire the same self-energy and one can write
$S_{\pi}(-q_0,\vec{q};T)=-S_{\pi}(q_0,\vec{q};T)$ with the
subsequent simplification of Eq.~(\ref{Lehmann}).}. Combining
Eqs.~(\ref{eq:prop1}) and (\ref{Lehmann}), conveniently continued
analytically from the Matsubara frequencies onto the real energy axis,
one can write
\begin{equation}
S_M(\omega,{\vec q}; T)= -\frac{1}{\pi} {\rm Im}\, D_M(\omega,{\vec q};T)
= -\frac{1}{\pi}\frac{{\rm Im}\, \Pi_M(\omega,\vec{q};T)}{\mid
\omega^2-\vec{q}\,^2-m_M^2- \Pi_M(\omega,\vec{q};T) \mid^2 } \ .
\label{eq:spec}
\end{equation}

Applying the finite-temperature Feynman rules, the meson-baryon loop function
in the ITF reads
\begin{eqnarray}
\label{G_ITF}
{\cal G}_{MB}(W_m,\vec{P};T) &=& - T \int \frac{d^3q}{(2\pi)^3} \,
\sum_n \frac{1}{{\rm i} W_m - {\rm i}\omega_n - E_B(\vec{P}-\vec{q},T)} \,
\nonumber \\
&\times&
\int_0^{\infty} d\omega \,
\left( \frac{S_M(\omega,\vec{q};T)}{{\rm i}\omega_n - \omega}
- \frac{S_{\bar M}(\omega,\vec{q};T)}{{\rm i}\omega_n + \omega} \right)
\,\,\, ,
\end{eqnarray}
where $\vec{P}$ is the external
total three-momentum and $W_m$ an external fermionic frequency,
${\rm i} W_m={\rm i} (2m+1)\pi T + \mu_B$.
Note that we follow a quasi-relativistic description of baryon fields all
throughout this work: in Eq.~(\ref{G_ITF}) the negative energy part of the
baryon propagator has been neglected but we use relativistic dispersion
relations.  All the Dirac structure is included in the definition of the tree
level amplitudes. The Matsubara sums can be performed using standard complex
analysis techniques for each of the two terms in the meson propagator and one
finds
\begin{eqnarray}
\label{G_ITF:Matsu-summed}
{\cal G}_{MB}(W_m,\vec{P};T) &=&
\int \frac{d^3q}{(2\pi)^3} \,
\int_0^{\infty} d\omega \,
\left[ S_M(\omega,\vec{q};T) \,
\frac{1-n_B(\vec{P}-\vec{q},T)+f(\omega,T)}
{{\rm i} W_m - \omega - E_B(\vec{P}-\vec{q},T)} \right.
\nonumber \\
&+&
\left.
S_{\bar M}(\omega,\vec{q};T) \,
\frac{n_B(\vec{P}-\vec{q},T)+f(\omega,T)}
{{\rm i} W_m + \omega - E_B(\vec{P}-\vec{q},T)} \, \right]
\,\,\, ,
\end{eqnarray}
with $f(\omega,T) = [\exp (\omega / T) - 1]^{-1}$ the meson Bose distribution
function at temperature $T$.
The former expression can be analytically continued onto the
real energy axis, $G_{MB}(P_0+{\rm i} \varepsilon \, ,\vec{P}; T) = {\cal
G}_{MB}({\rm i} W_m \to P_0 + {\rm i} \varepsilon \, , \vec{P}; T )$,
cf.~Eq.~(\ref{G_ITF:Matsu-summed}). With these medium modifications the
meson-baryon retarded propagator at finite temperature (and density) reads
\begin{eqnarray}
\label{eq:gmed}
{G}_{\bar K(K) N}(P_0+{\rm i} \varepsilon,\vec{P};T)
&=&\int \frac{d^3 q}{(2 \pi)^3}
\frac{M_N}{E_N (\vec{P}-\vec{q},T)} \nonumber \\
&\times  &\left[ \int_0^\infty d\omega
 S_{\bar K(K)}(\omega,{\vec q};T)
\frac{1-n_N(\vec{P}-\vec{q},T)}{P_0 + {\rm i} \varepsilon - \omega
- E_N
(\vec{P}-\vec{q},T) } \right. \nonumber \\
&+& \left. \int_0^\infty d\omega
 S_{K (\bar K)}(\omega,{\vec q};T)
\frac{n_N(\vec{P}-\vec{q},T)} {P_0 +{\rm i} \varepsilon + \omega -
E_N(\vec{P}-\vec{q},T)} \right] \ ,
\end{eqnarray}
for $\bar K(K)N$ states and
\begin{eqnarray}
\label{eq:gmed_piY}
{G}_{\pi Y}(P_0+{\rm i} \varepsilon,\vec{P}; T)
&= & \int \frac{d^3 q}{(2 \pi)^3} \frac{M_{Y}}{E_{Y}
(\vec{P}-\vec{q},T)} \nonumber \\
& \times &
\int_0^\infty d\omega
 S_\pi(\omega,{\vec q},T)
\left[
\frac{1+f(\omega,T)}
{P_0 + {\rm i} \varepsilon - \omega - E_{Y}
(\vec{P}-\vec{q},T) }   \right.
\nonumber \\
& + &
\left.
\frac{f(\omega,T)}
{P_0 + {\rm i} \varepsilon + \omega - E_{Y}
(\vec{P}-\vec{q},T) } \right]
\end{eqnarray}
for $\pi \Lambda$ or $\pi \Sigma$ states, where $P=(P_0,\vec{P})$ is the total
two-particle momentum and ${\vec q}$ is the meson three-momentum in the nuclear medium
rest frame.  Note that for consistency with the free space meson-baryon
propagator, given in Eq.~(\ref{G_vacuum}), we have included the normalization
factor $M_B / E_B$ in the baryon propagator. We have explicitly written the
temperature dependence of the baryon energies indicating that we account for
mean-field binding potentials as discussed above. The second term in the $\bar K
N$ loop function typically provides a small, real contribution for the energy
range in $P_0$ we are interested in. In order to simplify the numerical
evaluation of the self-consistent ${\bar K}N$ amplitude we replace
$S_{K}(\omega, \vec q;T )$ by a free-space delta function in
Eq.~(\ref{eq:gmed}). This approximation is sensible as long as the $K$ spectral
function in the medium still peaks at the quasi-particle energy and the latter
does not differ much from the energy in vacuum, as we will confirm in
Sect.~\ref{sec:Resul}. In Eq.~(\ref{eq:gmed}) we have neglected the kaon
distribution function, since we expect Bose enhancement to be relevant only for
the lightest meson species in the range of temperatures explored in the present
 study, $T = 0$ -- $150$~MeV.

The $\pi Y$ loop function, in particular, incorporates the $1+f(\omega ,T)$
enhancement factor which accounts for the  contribution from thermal pions at
finite temperature, cf.~Eq.~(\ref{eq:gmed_piY}).
In this case, we have neglected the fermion distribution for the participating
hyperons, which is a reasonable approximation for the range of temperature and
baryonic chemical potential that we have studied (cf.~Table~\ref{table:dmuB}).

In the case of $\eta \Lambda$, $\eta \Sigma$ and $K \Xi$ intermediate states,
we simply consider
the meson propagator in free space and include only the effective baryon
energies modified by the mean-field binding potential, namely
\begin{eqnarray}
G_l(P_0+{\rm i} \varepsilon,\vec{P};T)= \int \frac{d^3 q}{(2 \pi)^3} \,
\frac{1}{2 \omega_l (\vec q\,)} \frac{M_l}{E_l (\vec{P}-\vec{q},T)} \,
\frac{1}{P_0 +
{\rm i} \varepsilon - \omega_l (\vec{q}\,) - E_l (\vec{P}-\vec{q},T) } \, .
\label{eq:gprop}
\end{eqnarray}
The latter channels are less relevant in the unitarization procedure of the
$s$-wave scattering amplitude. They are important to maintain SU(3)
symmetry through using a complete basis of states in the coupled-channels 
procedure, as well
as for producing a better description of branching ratios between the various
scattering transitions at threshold. However, the width and position of the
$\Lambda (1405)$ are basically determined from the unitarized coupling of
${\bar K} N$ and $\pi \Sigma$ channels \cite{Oset:1997it}.
In addition, the changes that kaons \cite{Waas:1996fy,Tolos:2005jg} and
$\eta$ mesons \cite{Waas:1997pe,GarciaRecio:2002cu} experience
in the medium at moderate densities are comparably weaker than for $\pi$ and
${\bar K}$, which justifies the simplification adopted here.

As mentioned above, all meson-baryon loop functions in our approach are
regularized with a cut-off in the three-momentum integration, $q_{\rm max}$. We
adopt here the regularization scale that was set in \cite{Oset:1997it}, where
the ${\bar K}N$ scattering amplitude was evaluated in free space, leading to a
remarkable description of several ${\bar K} N$ scattering observables and the
dynamical generation of the $\Lambda (1405)$ resonance with a single parameter,
$q_{\rm max}=630$~MeV/c.

Finally, we obtain the in-medium $s$-wave $\bar K(K)$ self-energy  by
integrating  $T_{\bar K (K)  N}$ over the nucleon Fermi distribution at a given
temperature,
\begin{eqnarray}
\Pi^s_{\bar K(K)}(q_0,{\vec q};T)= \int \frac{d^3p}{(2\pi)^3}\,
n_N(\vec{p},T) \, [{T}^{(I=0)}_{\bar K(K)N}(P_0,\vec{P};T) +
3{T}^{(I=1)}_{\bar K(K)N}(P_0,\vec{P};T)]\ , \label{eq:selfd}
\end{eqnarray}
where $P_0=q_0+E_N(\vec{p},T)$ and $\vec{P}=\vec{q}+\vec{p}$ are
the total energy and momentum of the $\bar K(K)N$ pair in the
nuclear medium rest frame, and $q$ stands for the
momentum of the $\bar K(K)$ meson also in this frame. The kaon
self-energy must be determined self-consistently since it is
obtained from the in-medium amplitude, $ T_{\bar K(K)N}$, which
requires the evaluation of the $\bar K(K)N$ loop function,
$G_{\bar K(K)N}$, and the latter itself is a function of
$\Pi_{\bar K (K)}(q_0, \vec q; T)$ through the kaon spectral
function, cf.~Eqs.~(\ref{eq:spec}), (\ref{eq:gmed}).
Note that Eq.~(\ref{eq:selfd}) is valid in cold nuclear matter.
In the Appendix we
provide a derivation of the $s$-wave self-energy from the kaon
nucleon $T$-matrix in ITF and give arguments for the validity of
Eq.~(\ref{eq:selfd}) as a sensible approximation.

\subsection{$p$-wave kaon self-energy}
\label{ssec:pwave-kaon-self-energy}

The main contribution to the $p$-wave kaon self-energy comes from
the $\Lambda$ and $\Sigma$ pole terms, which are obtained from the
axial-vector couplings in the Lagrangian ($D$ and $F$ terms in
Eq.~(\ref{chiralLag})). The $\Sigma^* (1385)$ pole term is also
included explicitly with couplings to the kaon-nucleon states
which were evaluated from SU(6) symmetry arguments in
\cite{Oset:2000eg}.

In Ref.~\cite{TOL06}, the combined $s$- and $p$-wave $\bar K$
self-energy was obtained self-consistently in cold nuclear matter
by unitarizing the tree level amplitudes in both channels.
Unitarization of the $p$-wave channel, however, did not provide
dramatic effects over the tree level pole terms. Those were used
in \cite{Oset:2000eg} to evaluate the $p$-wave self-energy. The
latter is built up from  hyperon-hole ($Yh$) excitations and it
provides sizable strength in the $\bar K$ spectral function below
the quasi-particle peak and a moderate repulsion (as it is
expected from the excitation of subthreshold resonances) at
nuclear matter density. Medium effects and unitarization actually
move the position of the baryon pole with respect to that in free
space~\cite{TOL06}. We incorporate this behaviour here in an effective
way through the use of baryon mean field potentials and 
re-evaluate the relevant many-body diagrams for the
$p$-wave self-energy at finite temperature and density,
which considerably simplifies the numerical
task.
We shall
work in the ITF, improving on some approximations of former
evaluations \cite{Oset:2000eg,Cabrera:2002hc,Tolos:2002ud}. Hence,
the total $K$ and ${\bar K}$ self-energy will consist of the sum
of the $s$- and $p$-wave contributions described here and in the
former section. Note that  the $p$-wave self-energy enters the
self-consistent calculation of the $s$-wave self-energy as the $K$
and ${\bar K}$ meson propagators in the intermediate states on the
$s$-wave kaon-nucleon amplitude are dressed with the total
self-energy at each iteration.

We can write the ${\bar K}$ $p$-wave self-energy as the sum of
each of the $Yh$ contributions,
\begin{eqnarray}
\label{eq:pwave-expression}
\Pi_{\bar{K}}^p(q_0,\vec{q}; T) &=& \frac{1}{2} \,
\tilde{V}^2_{\bar K N \Lambda} \, \vec{q} \,^2 \,
f_{\Lambda}^2(q_0,\vec{q}\,) \, U_{\Lambda N^{-1}}(q_0,\vec{q}; T)
\nonumber \\
&+& \frac{3}{2} \, \tilde{V}^2_{\bar{K} N \Sigma} \, \vec{q}\, ^2
\, f_{\Sigma}^2(q_0,\vec{q}\,) \, U_{\Sigma N^{-1}}(q_0,\vec{q};
T)
\nonumber \\
&+& \frac{1}{2} \, \tilde{V}^2_{\bar K N \Sigma^*} \, \vec{q}\, ^2
f_{\Sigma^*}^2(q_0,\vec{q}\,) \, U_{\Sigma^* N^{-1}}(q_0,\vec{q};
T) \ , \label{eq:self}
\end{eqnarray}
where $U_{Y}$ stands for the $YN^{-1}$ Lindhard function at finite
temperature and baryonic density, and $\tilde{V}^2_{\bar K NY}$
represents the $\bar K NY$ coupling from the chiral Lagrangian in
a  non-relativistic approximation (leading order in a $M_B^{-1}$
expansion)  and includes the required isospin multiplicity. They
can be found, for instance, in Ref.~\cite{Oset:2000eg}. The $f_Y$
factors account for relativistic recoil corrections to the ${\bar
K}NY$ vertices, which improve on the lowest order approximation
and still allow to write the self-energy in a simple form, where
all the dynamical information from the $p$-wave coupling is
factorized out of the momentum sum in the fermionic loop. These
factors read
\begin{eqnarray}
\label{eq:recoilfactors}
f_{\Lambda , \Sigma}^2(q_0,\vec{q}\,) &=& 
\left[ 2 \,M_{\Lambda , \Sigma} + 2\,q_0\, (M_{\Lambda , \Sigma}-M_N)
- q^2 + 2 \, M_N E_{\Lambda , \Sigma} (\vec{q}\,) \right]
/ 4 \, M_N E_{\Lambda , \Sigma} (\vec{q}\,) \ \ \ ,
\nonumber \\
f_{\Sigma^*}^2(q_0,\vec{q}\,) &=&
(1 - q_0 /M_{\Sigma^*})^2 \ \ \ .
\end{eqnarray}
We have also accounted for the finite size of the vertices by
incorporating phenomenological hadronic form factors of dipole
type via the replacement $\vec{q}\,^2 \to F_K (\vec{q}\,^2) \,
\vec{q}\,^2$ with $F_K (\vec{q}\,^2) = (\Lambda_K^2 / (\Lambda_K^2
+ \vec{q}\,^2))^2$, where $\Lambda_K=1050$~MeV.
We provide explicit expressions for $U_{Y N^{-1}}$ in Appendix~\ref{app-Linds}.

\begin{figure}[t]
\begin{center}
\includegraphics[width=14cm]{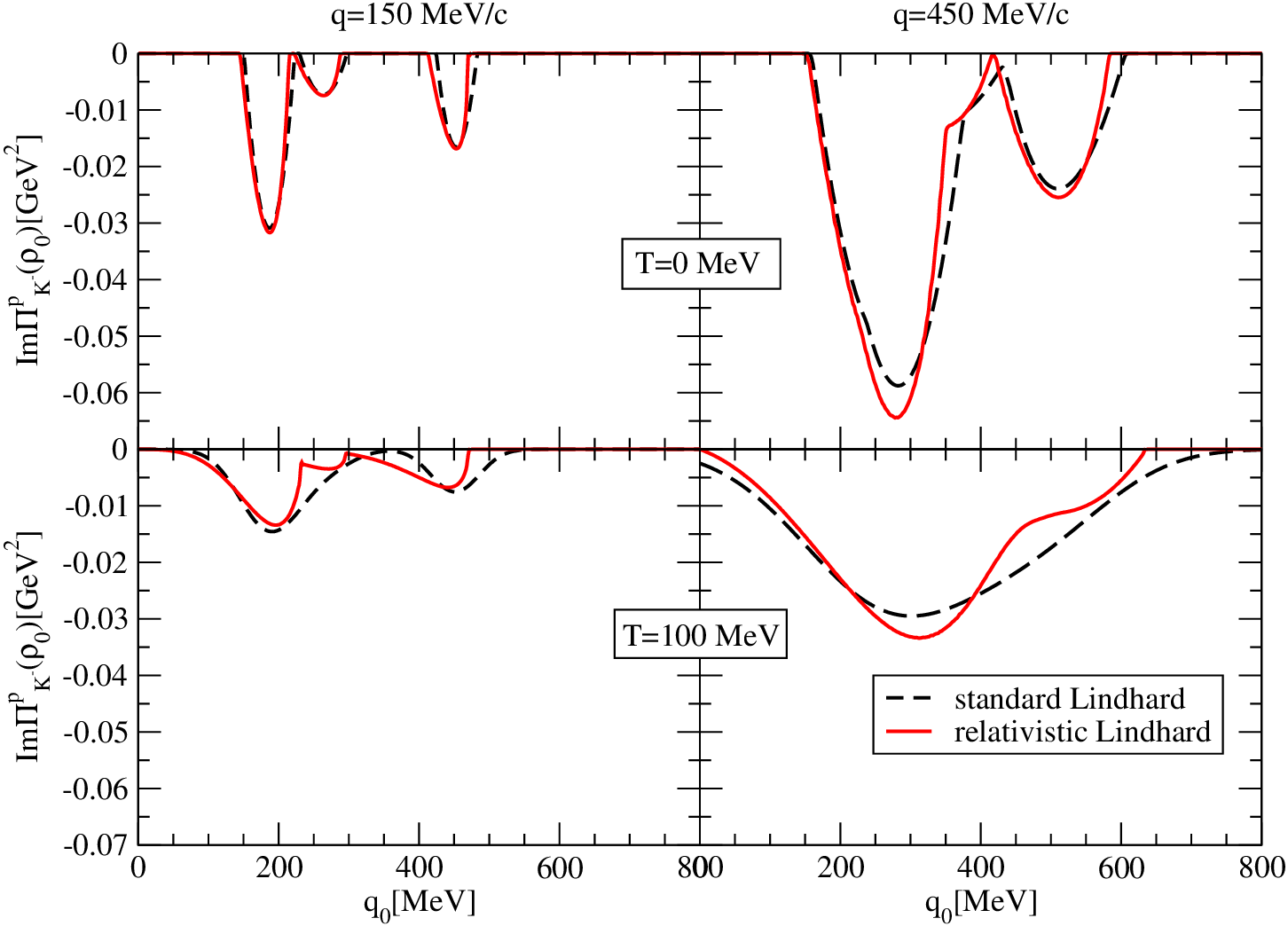}
\caption{Imaginary part of the $\bar K$ $p$-wave self-energy from different
evaluations of the $YN^{-1}$ Lindhard function (see text) at $\rho=\rho_0$. The
upper panels correspond to $T=0$, whereas in the
lower panels $T=100$~MeV.}
\label{fig_pwave_comparison}
\end{center}
\end{figure}

In Refs.~\cite{Oset:2000eg,Cabrera:2002hc} the $K$ and ${\bar K}$
self-energy in $p$-wave was obtained for cold nuclear matter, and
it was extended to the finite temperature case in
\cite{Tolos:2002ud}. We would like to present here a comparison of
our results for different approximations as the ones used in
former evaluations, particularly in the $T \to 0$  limit. In
Fig.~\ref{fig_pwave_comparison} we present the imaginary part of
the ${\bar K}$ $p$-wave self-energy as a function of the kaon
energy, evaluated at nuclear matter density and two different kaon
momenta. In the upper panels the dashed lines have been obtained
by using, in Eq.~(\ref{eq:self}), the standard non-relativistic evaluation
of the $YN^{-1}$ Lindhard function
\cite{Oset:2000eg,Cabrera:2002hc}, the analytical expression of
which is also quoted in Appendix~\ref{app-Linds}.
The solid lines correspond to
our calculation in Eq.~(\ref{eq:Lind-rel-YN-elaborate}) for
$T=0$~MeV. Both results agree quite well which ensures that our
calculation has the correct $T\to 0$ limit. The observed
differences, particularly the threshold energies at which each
$YN^{-1}$ component is open/closed, are related to the use of
non-relativistic baryon energies (in dashed lines) and some
further approximations that we discuss below.

At $T=100$~MeV (lower panels), the dashed lines correspond to the
evaluation of the Lindhard function in \cite{Tolos:2002ud}, which
extends the zero-temperature, non-relativistic calculation
by replacing the nucleon occupation number,
$\theta(p_F-p)$, by the corresponding Fermi-Dirac distribution,
$n_N(\vec{p},T)$. Note that, in our result (solid lines), the
Matsubara sum automatically generates a non-vanishing term
proportional to the hyperon distribution function,
cf.~Eqs.~(\ref{eq:Lind-rel-YN},\ref{eq:Lind-rel-YN-elaborate}) in
Appendix~\ref{app-Linds}.
Despite the absence of a crossed kinematics mechanism (as for
instance in $ph$ and $\Delta h$ excitations for pions) this
guarantees that the imaginary part of the $\bar K$ self-energy
identically vanishes at zero energy ($q_0=0$) as it follows from
the (retarded) crossing property of the thermal self-energy, ${\rm
Im} \, \Pi_{\bar K}^p (-q_0,\vec{q};T) = - {\rm Im} \, \Pi_K^p
(q_0,\vec{q};T)$ \footnote{At zero energy ($q_0=0$) the $K$ and
$\bar K$ modes cannot be distinguished and thus ${\rm Im} \,
\Pi_{\bar K}^p (0,\vec{q};T) = - {\rm Im} \, \Pi_K^p (0,\vec{q};T)
\equiv 0$. This is accomplished exactly in
Eq.~(\ref{eq:Lind-rel-YN-elaborate}) in Appendix~\ref{app-Linds} for $q_0=0$.}. The two
calculations at $T=100$~MeV exhibit some non-trivial differences.
In the finite-$T$ extension of the non-relativistic result the
$YN^{-1}$ structures and thresholds are more diluted and the
$\Sigma$ component is not resolved even at low momentum, whereas
in the relativistic result the three components are clearly
identifiable at $q=150$~MeV$/c$. In both calculations the strength
of the imaginary part extends to lower energies so that the energy
gap in cold nuclear matter is absent here. However, as mentioned
before, the relativistic  calculation in the ITF vanishes exactly
at $q_0=0$ whereas the non-relativistic, finite-$T$ extended
result does not. In addition, the standard non-relativistic
calculations (at $T=0$ and finite $T$)
\cite{Oset:2000eg,Cabrera:2002hc,Tolos:2002ud} omit $\vec{p}\,^2$
terms proportional  to $(M_Y^{-1}-M_N^{-1})$ in the energy balance
of the two-particle propagator, which are responsible for cutting
the available $YN^{-1}$ phase space for high external energies
($q_0$). As a consequence, the dashed lines in the lower panels
exhibit a high energy tail which is attenuated by the nucleon
distribution whereas the relativistic result has a clear
(temperature dependent) end point. This happens for each of the
excited hyperon components independently and it is particularly
visible for the $\Sigma^*$. The $\Lambda$ and $\Sigma$ tails mix
with each other and are responsible for the washing out of the
$\Sigma$ structure at finite momentum.

Finally, the $K$ $p$-wave self-energy can be obtained following the same
procedure as above. The excitation mechanisms in this case correspond exactly to
the crossed kinematics of those of the $\bar K$. Taking advantage of the
crossing property of the thermal self-energy, we obtain the $K$ self-energy from
the $\bar K$ one replacing $q_0 \to - q_0$ in $\Pi_{\bar K}^p (q_0,\vec{q};T)$
(modulo a sign flip in the imaginary part). The crossed kinematics causes the
$YN^{-1}$ excitations to be far off-shell. In cold nuclear matter, for instance,
$\Pi_K^p$ is real and mildly attractive. At finite temperature, though, the
fermion distributions of the nucleon and hyperon can accommodate low-energy
(off-shell) kaons and $\Pi_K^p$ receives a small imaginary part which rapidly
decays with increasing energy.

\subsection{Pion self-energy}
\label{ssec:pion-self-energy}

We briefly discuss here the relevant many-body mechanisms that
modify the pion propagator, which enters the evaluation of the
in-medium ${\bar K} N$ amplitude, cf.~Eqs.~(\ref{eq:BSalgeb}),
(\ref{eq:gmed_piY}). In cold nuclear matter, the pion spectral
function exhibits a mixture of the pion quasi-particle mode and
$ph$, $\Delta h$ excitations \cite{Oset:1989ey}. The meson-baryon
chiral Lagrangian in Eq.~(\ref{chiralLag}) provides the $\pi NN$
$p$-wave vertex, while the $\pi N\Delta$ vertex can be determined
from the standard non-relativistic derivation of the
Raritta-Schwinger interaction Lagrangian. However, we shall use
phenomenological $\pi NN$ and $\pi N \Delta$ coupling constants
determined from analysis of pion nucleon and pion nucleus
reactions. Their values are $f_N/m_\pi = 0.007244$~MeV$^{-1}$
and $f_{\Delta}/f_N=2.13$. The lowest order $p$-wave pion self-energy
due to $ph$ and $\Delta h$ excitations then reads
\begin{equation}
\label{eq:piself-ph-Dh}
\Pi_{\pi NN^{-1}+\pi\Delta N^{-1}}^p (q_0,\vec{q};T) =
\left( \frac{f_N}{m_{\pi}} \right) ^2
\vec{q}\,^2 \, \left[ U_{NN^{-1}} (q_0,\vec{q};T)
+ U_{\Delta N^{-1}} (q_0,\vec{q};T) \right]
\,\,\, ,
\end{equation}
where the finite temperature Lindhard functions for the $ph$ and
$\Delta h$ excitations are given in Appendix~\ref{app-Linds}. Note that, for
convenience, we have absorbed the $\pi N \Delta$ coupling in the
definition of $U_{\Delta N^{-1}}$.

The strength of the considered collective modes is modified by
repulsive, spin-isospin $NN$ and $N\Delta$ short range
correlations \cite{Oset:1981ih}, which we include in a
phenomenological way with a single Landau-Migdal interaction
parameter, $g'=0.7$. The RPA-summed pion self-energy then reads
\begin{equation}
\label{eq:piself-total}
\Pi^p_{\pi} (q_0,\vec{q};T) =
\frac{\left( \frac{f_N}{m_{\pi}} \right) ^2
F_{\pi}(\vec{q}\,^2) \, \vec{q}\,^2 \,
\left[ U_{NN^{-1}} (q_0,\vec{q};T) + U_{\Delta N^{-1}} (q_0,\vec{q};T) \right]}
{1 - \left( \frac{f_N}{m_{\pi}} \right) ^2 \, g' \,
\left[ U_{NN^{-1}} (q_0,\vec{q};T) + U_{\Delta N^{-1}} (q_0,\vec{q};T) \right]}
\,\,\, ,
\end{equation}
which also contains the effect of the same monopole form factor at
each $\pi NN$ and $\pi N \Delta$ vertex as used in $T=0$ studies, 
namely $F_{\pi}(\vec{q}\,^2) = (\Lambda_{\pi}^2 - m_{\pi}^2) /
[\Lambda_{\pi}^2 - (q_0)^2 + \vec{q}\,^2 ]$, with
$\Lambda_{\pi}=1200$~MeV, as is needed in the empirical study of $NN$
interactions.
Finally, for consistence with former
evaluations of the pion self-energy, we have also accounted for
one-body $s$-wave scattering and $2p2h$ mechanisms, following the
results in Refs.~\cite{Ramos:1994xy,Seki:1983sh,Meirav:1988pn},
which we have kept to be the same as in $T=0$.

\begin{figure}[t]
\begin{center}
\includegraphics[width=14cm]{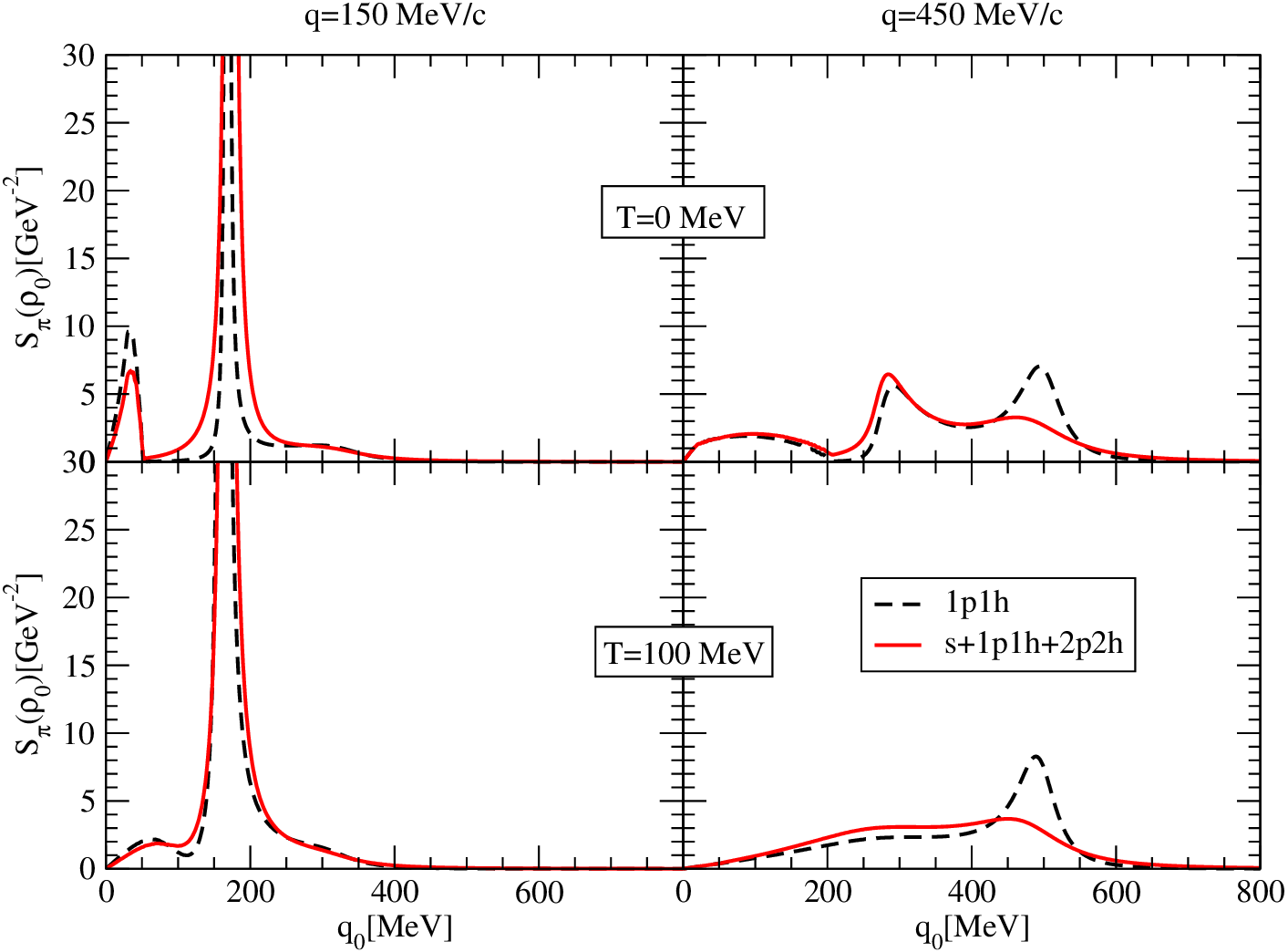}
\caption{Pion spectral function at $\rho=\rho_0$ for two different momenta and
temperatures, $T=0$ (upper panels) and $T=100$~MeV (lower panels). The dashed
lines correspond to the $p$-wave self-energy calculation including $ph$, $\Delta
h$ and short range correlations. The solid lines include, in addition, the
(small) $s$-wave self-energy and the $2p2h$ absorption mechanisms.}
\label{fig:pion-spectral}
\end{center}
\end{figure}

In Fig.~\ref{fig:pion-spectral} we show the pion spectral function
at normal nuclear matter density for two different momenta. At
$T=0$ (upper panels) one can easily distinguish the different
modes populating the spectral function. At low momentum, the pion
quasi-particle peak carries most of the strength together with the
$ph$ structure at lower energies. The $\Delta h$ mode starts to
manifest to the right hand side of the pion quasi-particle peak.
Note that the pion mode feels a sizable attraction with respect to
that in free space. At higher momentum, the excitation of the
$\Delta$ is clearly visible and provides a considerable amount of
strength which mixes with the pion mode. As a consequence, the
latter broadens considerably. The solid lines include also the
contributions from the pion $s$-wave self-energy and two-body
absorption. These mechanisms, especially the latter, generate a
background of strength which further broadens the spectral
function, softening in particular the pion peak at low momentum
and the $\Delta$ excitation at higher momentum.

The lower panels correspond to a temperature of $T=100$~MeV. The
softening of the nucleon occupation number due to thermal motion
causes a broadening of the three modes present in the spectral
function. At $q=450$~MeV, the $ph$, $\Delta h$ and pion peaks are
completely mixed, although some distinctive strength still
prevails at the $\Delta h$ excitation energy.
The $s$-wave and $2p2h$ self-energy terms completely wash out the
structures that were still visible at zero temperature. For higher
temperatures we have checked that no further structures can be
resolved in the spectral function, in agreement with
Ref.~\cite{Rapp-rho-finiteT}. We however find differences at the
numerical level, as we expected, since we have implemented
different hadronic form factors and Landau-Migdal interactions in
our model. Moreover, to keep closer to phenomenology, we have also
implemented in our model the energy-dependent $p$-wave decay width
of the $\Delta$, which favors the mixing of the $\Delta h$
excitation with the pion mode.

\section{Results and Discussion}
\label{sec:Resul}

\subsection{The $\bar K$ meson spectral function in a hot nuclear medium}
\label{ssec:Resul-Kbar-spectral}

\begin{figure}[t]
\begin{center}
\includegraphics[width=14cm]{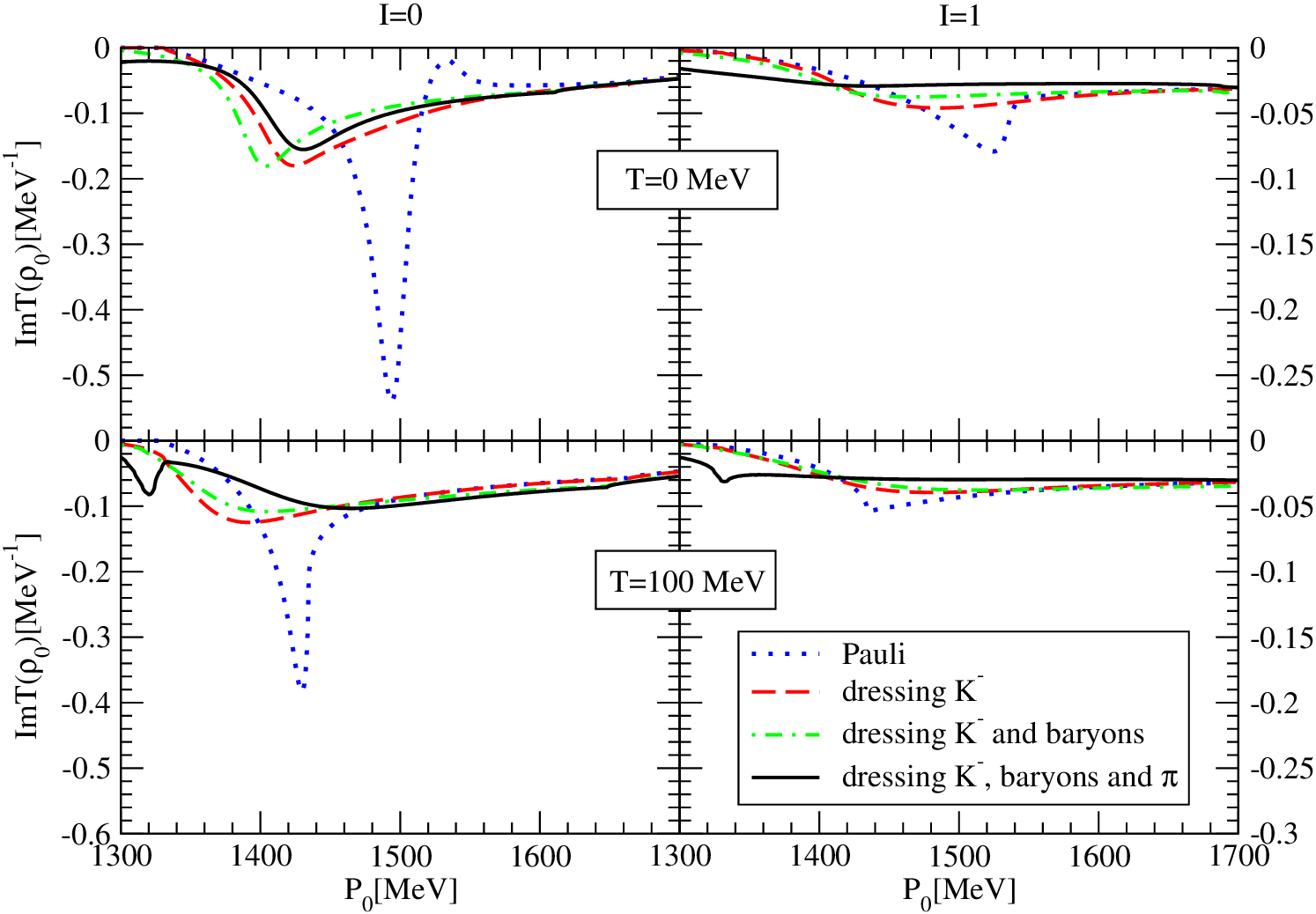}
\caption{Imaginary part of the in-medium ${\bar K} N$ $s$-wave amplitude for
$I=0$ and $I=1$ at $\rho_0$ as a function of the center-of-mass energy $P_0$,
for $T=0$, $100$~MeV, and for the different approaches discussed in the
text.}
\label{fig_amp}
\end{center}
\end{figure}

We start this section by showing in Fig.~\ref{fig_amp} the $s$-wave $\bar K N$
amplitude for $I=0$ and $I=1$ as a function of the center-of-mass energy, $P_0$,
calculated at nuclear matter density, $\rho_0$, and temperatures $T=0$ MeV 
(first row) and $T=100$ MeV (second row). We have considered four different
in-medium approaches: a first iteration that only includes Pauli blocking on the
nucleon intermediate states (dotted lines), and three other self-consistent
calculations of the $\bar K$ meson self-energy increasing gradually the degree
of complexity: one includes the dressing of the $\bar K$ meson (long-dashed
lines), another considers in addition the mean-field binding of the baryons in
the various intermediate states (dot-dashed lines), and, finally, the complete
model which includes also the pion self-energy (solid lines). Recall that the 
$I=0$ amplitude is governed by the behavior of the dynamically generated
$\Lambda(1405)$ resonance.

We start commenting on the $T=0$ results shown in the upper panels, where we
clearly see that the inclusion of Pauli blocking on the intermediate nucleon
states generates the $I=0$ $\Lambda(1405)$ at higher energies than its position
in free space. This has been discussed extensively in the literature
\cite{Lutz,Koch,Ramos:1999ku,Tolos:2002ud,TOL06,TOL00} and it is due to the 
restriction of available phase space in the unitarization procedure. The
self-consistent incorporation of the attractive ${\bar K}$ self-energy moves the
$\Lambda(1405)$ back in energy, closer to the free position, while it gets
diluted due to the opening of  the $\Lambda(1405) N \rightarrow \pi N \Lambda,
\pi N \Sigma$ decay modes \cite{Lutz,Ramos:1999ku,Tolos:2002ud,TOL06}. The
inclusion of baryon binding has mild effects, lowering slightly the position of
the resonance peak. A similar smoothing behavior is observed for the $I=1$
amplitude as we include medium modifications on the intermediate meson-baryon
states. As already pointed out in Ref.~\cite{Ramos:1999ku,Tolos:2002ud,TOL06},
when pions are dressed new channels are available, such as $\Lambda N N^{-1}$ or
$\Sigma N N^{-1}$ (and similarly with $\Delta N^{-1}$ components), so the
$\Lambda(1405)$ gets further diluted.

At a finite temperature of 100 MeV (lower panels), the  $\Lambda(1405)$
resonance gets diluted and is produced, in general,  at lower energies due to
the smearing of the Fermi surface that reduces the Pauli blocking effects.
When pions are dressed the strongly diluted resonance moves slightly to higher
energies compared to the zero temperature case. The cusp like structures that
appear at the lower energy side signal the opening of the $\pi\Sigma$ threshold
on top of the already opened $YNN^{-1}$ one.  Note that the cusp-like structure
appears enhanced  in the $I=0$ case. We believe that this is a manifestation  at
finite temperature and density of the two-pole structure of the $\Lambda(1405)$.
As seen in
Refs.~\cite{Oller:2000fj,Garcia-Recio:2002td,Jido:2002yz,Jido:2003cb,Garcia-Recio:2003ks},
this resonance is, in fact, the combination of two poles in the complex plane
that appear close in energy and couple strongly  to either $\bar K N$ or $\pi
\Sigma$ states. These two poles move apart and can be even resolved in a hot
medium because  density and temperature influence each of them differently.
Although not shown in the figure, we find that, at twice nuclear matter density,
the pole that couples more strongly to $\pi\Sigma$ states moves further below
the $\pi\Sigma$ threshold and acquires a clear Breit-Wigner shape. This
allows us to conclude that the cusp observed in the $I=0$ amplitude at $T=100$
MeV and $\rho=\rho_0$ is essentially a reflection of this pole. The shape of
the  corresponding resonance appears distorted with respect to a usual
Breit-Wigner because the pole  lies just below the threshold of the $\pi\Sigma$
channel to which it couples  very strongly, a behavior known as Flatt\'e effect
\cite{Flatte:1976xu}. Note that these structures are not seen in the $T=0$
results because the $\pi\Sigma$ threshold in that case is located at around 1295
MeV, out of the  range of the plot.

\begin{figure}[t]
\begin{center}
\includegraphics[width=14cm]{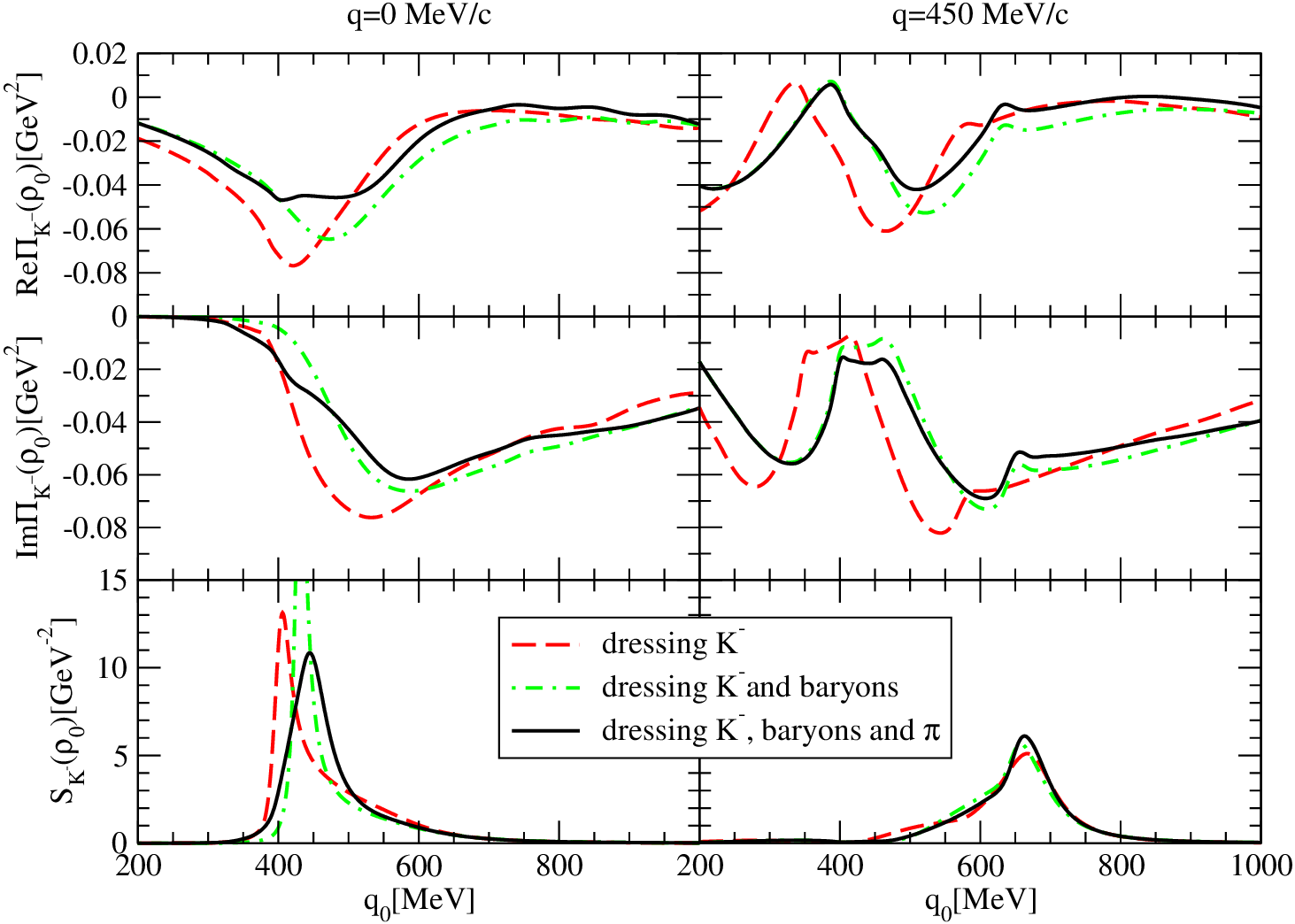}
\caption{
Real and imaginary parts of the $\bar K$ self-energy and spectral density,
as functions of the $\bar K$ energy,
for $q=0$ MeV/c and $q=450$ MeV/c. Results have been obtained at 
$T=0$ and $\rho_0$, including $(s+p)$-wave contributions, for the three
self-consistent approaches discussed in the text.
}
\label{fig_selftot_T0}
\end{center}
\end{figure}

Results for the ${\bar K}$
self-energy and spectral function at $\rho_0$ and $T=0$, including $s$- and 
$p$-wave contributions, obtained in the different
self-consistent approaches are compared in
Fig.~\ref{fig_selftot_T0}. At $q=0$ MeV/c, the inclusion of the baryon binding potential
(dot-dashed lines) moves the quasi-particle peak of the spectral function, which
is defined as
\begin{equation}
E_{qp}(\vec{q}\,)^2=\vec{q}\,^2+m_{\bar K}^2+{\rm
Re}\,\Pi_{\bar K}(E_{qp}(\vec{q}\,),\vec{q}\,) \ , \label{eq:Qparticle}
\end{equation}
to higher energies with respect to the case with no binding (dashed lines).
The pion dressing (solid lines) further alters the
behavior of the self-energy and, hence, that of the spectral function. The
attraction of the antikaon mode decreases, while its width increases due to the
opening of new decay channels induced by the $ph$ and $\Delta h$ pion
excitations.
At finite momentum, the $p$-wave $\Lambda N^{-1}$, $\Sigma N^{-1}$ and
$\Sigma^* N^{-1}$ excitation modes are clearly visible
 around 300, 400 and 600~MeV, respectively. The latter mode mixes
very strongly with the quasi-particle peak, making 
the differences between various
self-consistent approaches to be less visible in the ${\bar K}$ spectral function.
Similar results were obtained in Ref.~\cite{TOL06}. The differences between 
both $T=0$
calculations arise mainly by the use in this work of different (and more
realistic) baryon binding potentials.
The effects of the particular details of the nucleon spectrum on the $\bar K$
spectral function have been noted recently in the $T=0$ study of
Ref.~\cite{Lutz:2007bh}.

\begin{figure}[t]
\begin{center}
\includegraphics[width=14cm]{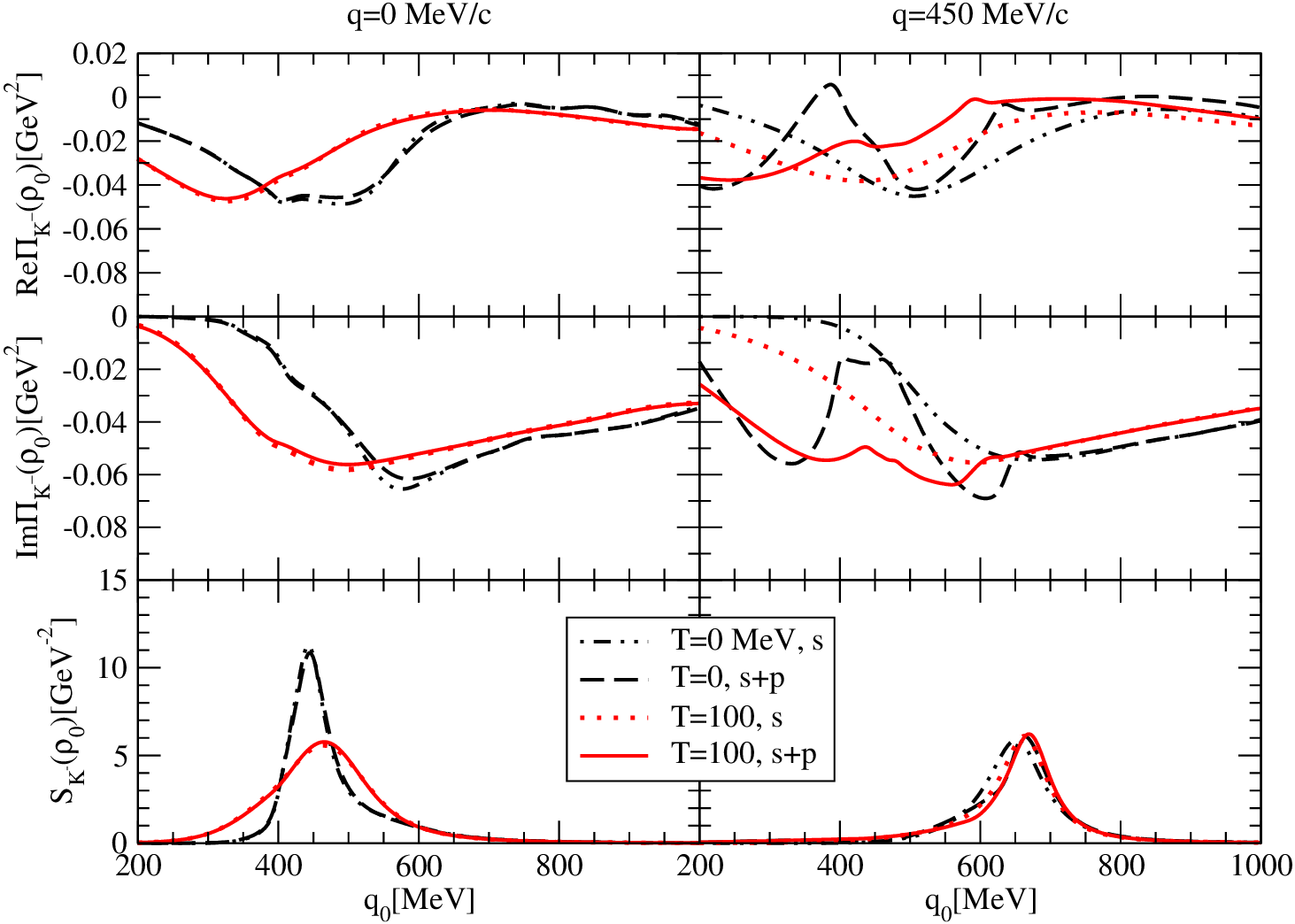}
\caption{Real and imaginary parts of the $\bar K$ self-energy and
spectral function at $\rho_0$, for $q=0,450$~MeV/c and $T=0$, $100$~MeV
as function of the $\bar K$ energy, including $s$- and $(s+p)$-wave
contributions in the full self-consistent calculation.
}
\label{fig_selftot}
\end{center}
\end{figure}

The effect of finite temperature on the different partial-wave contributions to
the $\bar K$ self-energy is shown in Fig.~\ref{fig_selftot}. In this figure we
display the real and imaginary parts of the $\bar K$ self-energy together with
the $\bar K$ spectral function for the self-consistent approximation that
dresses baryons and includes the pion self-energy. We show the results as a
function of the $\bar K$ energy for two different momenta, $q=0$~MeV/c (left
column) and $q=450$~MeV/c (right column). The different curves correspond to
$T=0$ and $T=100$ MeV including the $s$-wave and the $(s+p)$-wave
contributions.

According to Eq.~(\ref{eq:self}), the present model does not give an
explicit $p$-wave contribution to the ${\bar K}$ self-energy  
at zero momentum. The small differences between the $s$ and the $s+p$ 
calculations observed at $q=0$ MeV/c are due to the indirect effects of having 
included
the $p$-wave self-energy in the intermediate meson-baryon loop. 
The importance of the $p$-wave
self-energy is more evident at a finite momentum of $q=450$~MeV/c. 
The effect of the
subthreshold $\Lambda N^{-1}$, $\Sigma N^{-1}$ and $\Sigma^*
N^{-1}$  excitations is repulsive at the $\bar K N$ threshold. This
repulsion together with the strength below threshold induced by the those
excitations can be easily seen in the spectral function at finite momentum
(third row). The quasi-particle peak moves to higher energies while the spectral
function falls off slowly on the left-hand side.

Temperature results in a softening of the real and imaginary part of the
self-energy as the Fermi surface is smeared out. The peak of the spectral
function moves closer to the free position while it extends over a wider range
of energies.

\begin{figure}[ht]
\begin{center}
\includegraphics[width=14cm]{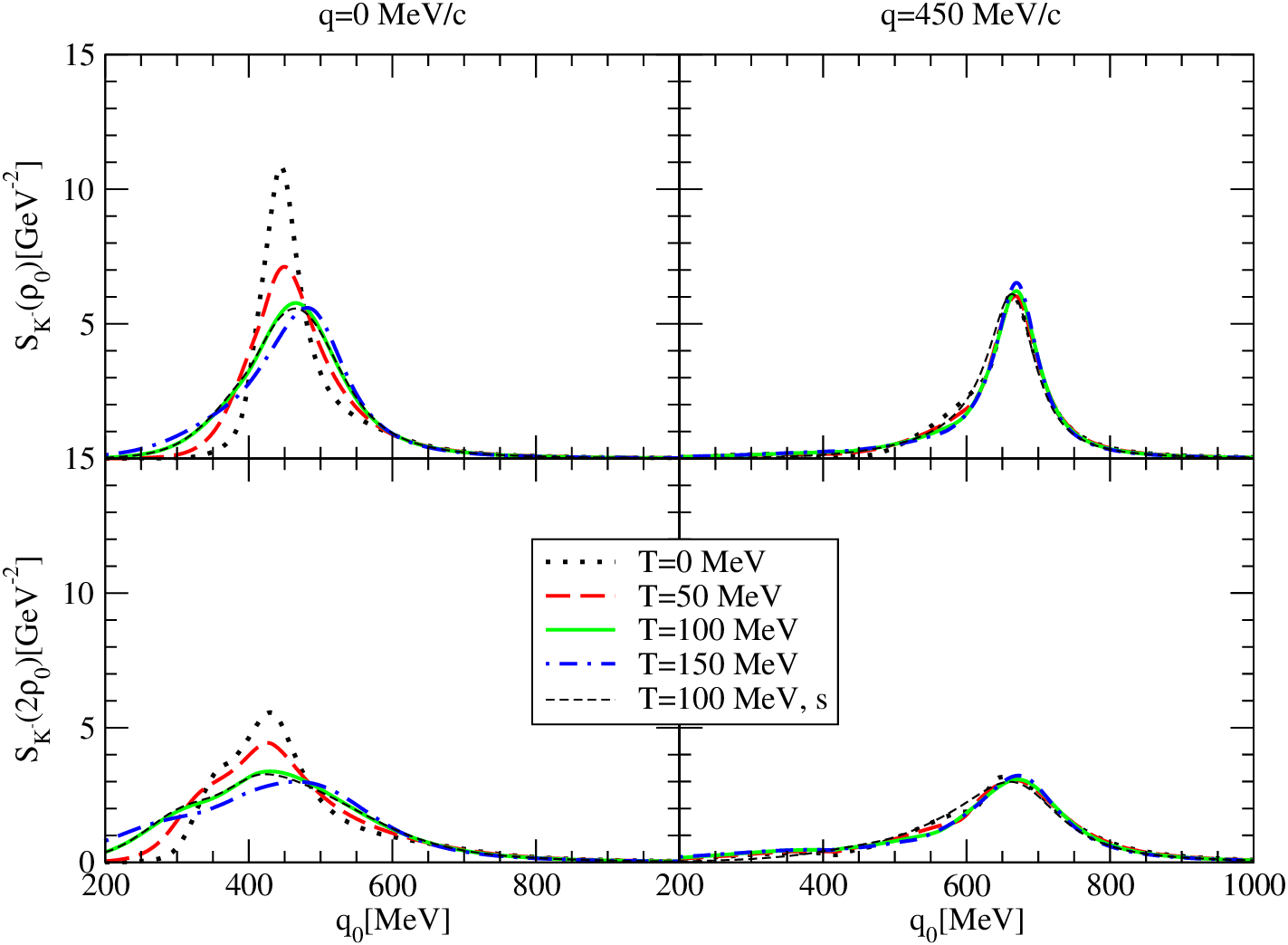}
\caption{The $\bar K$ meson spectral function for $q=0$ MeV/c and $q=450$ MeV/c at $\rho_0$ and $2\rho_0$  as a function of
the $\bar K$ meson energy for different temperatures and for the self-consistent
calculation including the dressing of baryons and pions.}
\label{fig_spectot_Kbar}
\end{center}
\end{figure}

For completeness, we show in  Fig.~\ref{fig_spectot_Kbar}  the evolution of the
$\bar K$ spectral function with increasing temperature for two different
densities, $\rho_0$ (upper row) and $2\rho_0$ (lower row), and two momenta,
$q=0$ MeV/c (left column) and $q=450$~MeV/c (right column), in the case of the
full self-consistent calculation which includes the dressing of baryons and the
pion self-energy.  At $q=0$ MeV/c the quasi-particle peak moves to higher energies
with increasing temperature due to the loss of strength of the attractive
effective ${\bar K}N$ interaction. Furthermore, the quasi-particle peak 
enhances its collisional broadening and gets mixed with the strength 
associated to the $\Lambda(1405)$ appearing both to the right (slow fall out)
and left (cusp-like structures) of the peak. All
these effects are less pronounced at a finite momentum of $q=450$ MeV/c. In this
case, the region of the quasi-particle peak is exploring ${\bar K}$ energies
of around 700~MeV, where the self-energy has a weaker energy dependence with
temperature, as seen in Fig.~\ref{fig_selftot}. We note that, as opposed to the
zero momentum case, the width of the
quasi-particle peak decreases with increasing temperature because of the
reduction of the inter-mixing 
with the $\Sigma^* N^{-1}$ excitations which get diluted in a hot medium.
As for the density effects, we just note that the quasi-particle peaks widens 
for larger nuclear density due to the enhancement of collision and absorption 
processes. In fact, a significant amount of strength is visible at energy values
substantially below the quasi-particle peak.
The fact that the $\bar{K}$ spectral function spreads to lower energies, even at
finite momentum, may have relevant implications on the phenomenology of the $\phi$
meson propagation and decay in a nuclear medium. We will further elaborate on
this point in the Conclusions section.

\subsection{$K$ meson in nuclear matter at finite temperature}
\label{ssec:Resul-Kaon-spectral}

\begin{figure}[ht]
\begin{center}
\includegraphics[width=14cm]{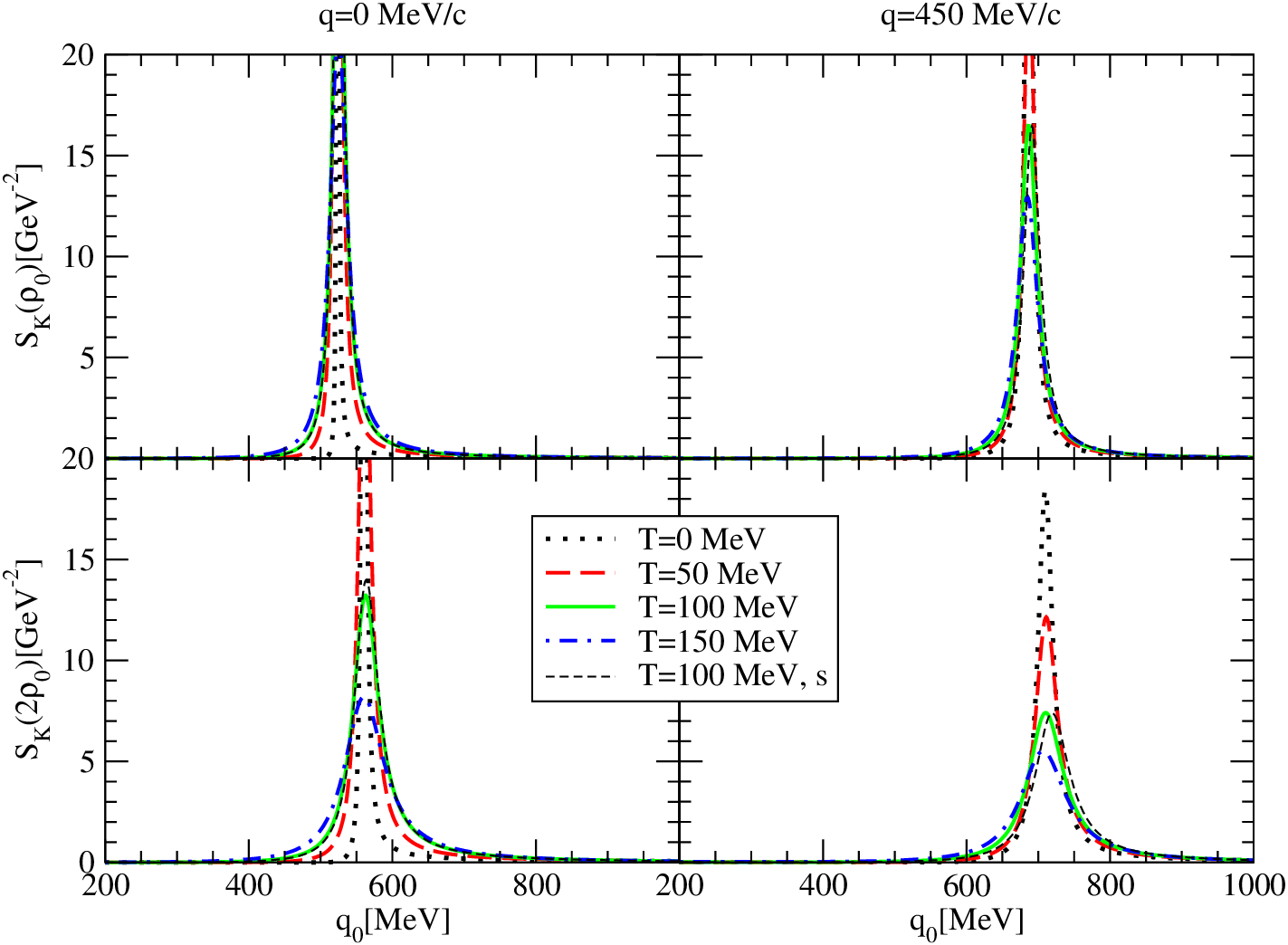}
\caption{The $K$ meson spectral function for $q=0$ MeV/c and $q=450$ MeV/c at
$\rho_0$ and $2\rho_0$  as a function of the $K$ meson energy for different
temperatures.}
\label{fig_spectot_K}
\end{center}
\end{figure}

The evolution with temperature and density of the properties of kaons is also a
matter of high interest. In particular, kaons are ideal probes to test the
high-density phase of relativistic heavy-ion collisions at incident energies
ranging from 0.6A to 2AGeV and for studying the stiffness of the nuclear
equation of state \cite{Forster:2007qk}. Moreover, there is a strong
interconnection between the $K^+$, $K^-$ and $\phi$ channels, which can lead to
important changes in the $\phi$-meson production in heavy-ion collisions
\cite{Mangiarotti:2003es}. 

In the $S=1$ sector, only the $KN$ channel is available and, hence, the
many-body dynamics of the $K$ meson in nuclear medium is simplified with respect
to the $\bar K N$ case. 
The $K$ spectral function at $q=0$ and
$q=450$  MeV/c is displayed in Fig.~\ref{fig_spectot_K} for the self-consistent
calculation that includes the dressing of baryons, for $\rho_0$ (upper row) and
$2\,\rho_0$ (lower row) and different temperatures.
The $K$ meson is described by a narrow quasi-particle peak which dilutes 
with temperature and density as the phase space for 
collisional $KN$ states increases. The $s$-wave
self-energy provides a moderate repulsion at the quasi-particle
energy, which translates into a shift of the $K$ spectral function to
higher energies with increasing density. In contrast to the $\bar K N$ case,
the inclusion of $p$-waves has a mild effect on the kaon
self-energy (compare thin-dashed lines to solid lines at $T=100$ MeV and $q=450$
MeV/c), as they arise from far off-shell $YN^{-1}$ excitations in crossed
kinematics. These excitations provide a small, attractive and barely energy
dependent contribution to the $K$ self-energy.

\begin{figure}[ht]
\begin{center}
\includegraphics[width=10cm]{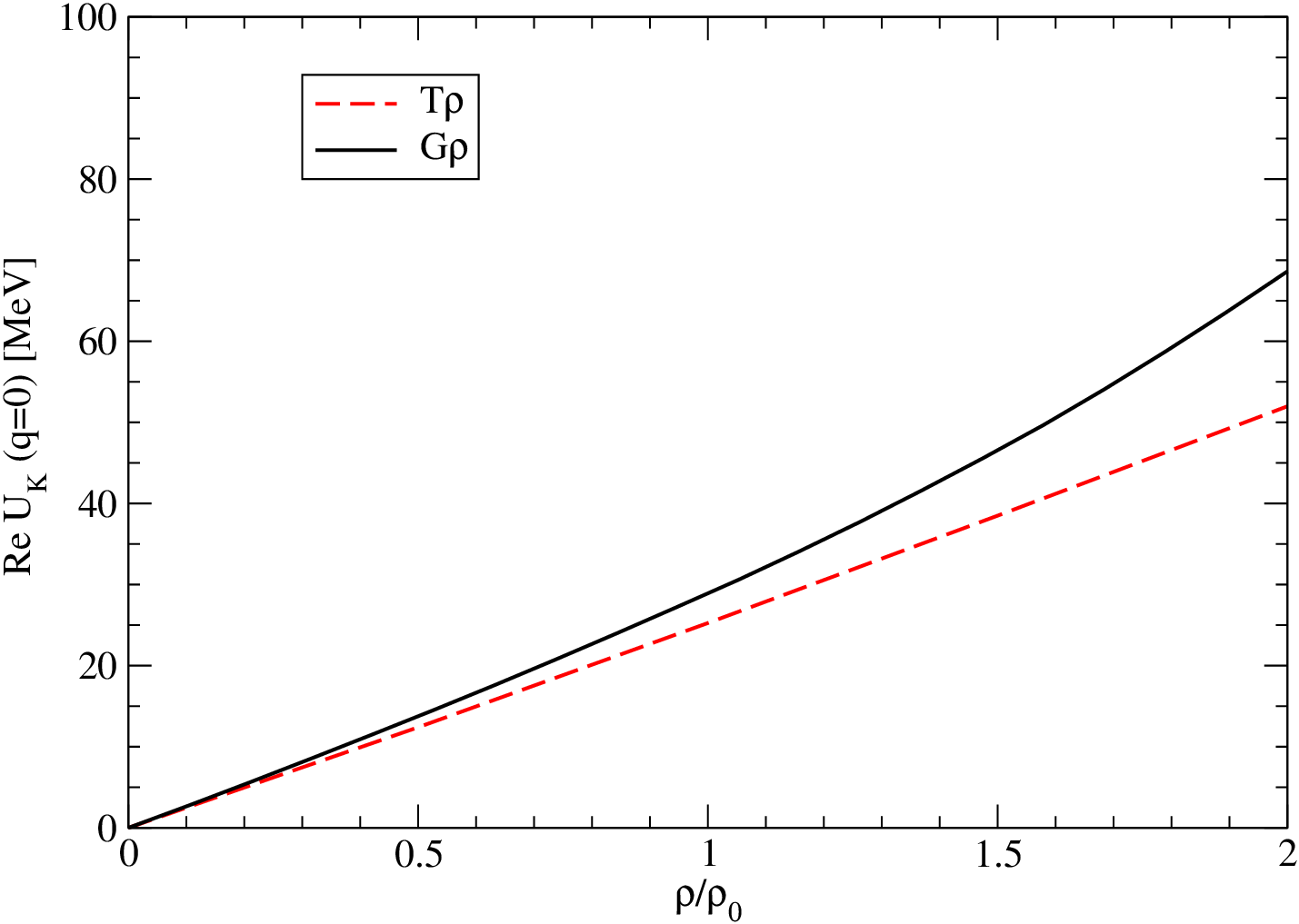}
\caption{The $K$ mass shift for $T=0$ MeV as a function of density, obtained
within the self-consistent calculation and in the $T\rho$ approximation.}
\label{fig_upot_dens}
\end{center}
\end{figure}

We can define the $K$ optical potential in the nuclear medium
as 
\begin{equation}
U_{K}(\vec{q},T)=\frac{\Pi_{K}(E_{qp}(\vec{q}\,),\vec{q},T)}{2\sqrt{m_{K}^2+\vec{q}\,^2}} \ ,
\label{eq:Kpot}
\end{equation}
which, at zero momentum, can be identified as the in-medium shift of the $K$
meson mass. The $K$ mass shift obtained in the self-consistent
calculation that considers also the nucleon binding effects is displayed in
Fig.~\ref{fig_upot_dens} as a function of the nuclear density for $T=0$ MeV. Our
self-consistent results are compared to those 
of the low-density or $T \rho$ approximation, obtained  
by replacing the medium-dependent amplitude by
the free-space one in Eq.~(\ref{eq:selfd}). We observe that the kaon potential
at nuclear saturation density in the $T \rho$ approximation is 4 MeV less
repulsive that in the case of the self-consistent approach, which gives a
repulsion of 29 MeV. This value is in qualitative agreement
with other self-consistent calculations  \cite{Tolos:2005jg,LUTZ-KORPA} and
close to the 20 MeV of repulsion
obtained in $K^+$ production on nuclei by the ANKE experiment of the COSY
collaboration \cite{Nekipelov}. We conclude that the
low-density theorem for
densities below normal nuclear matter is fulfilled within 15 \% due to the smooth energy
dependence of the $K N$ interaction tied to the absence of resonant states 
close to the $KN$ threshold.

\subsection{In-medium $\bar K$ and $K$ optical potentials at finite
temperature}
\label{ssec:optical}

\begin{figure}[ht]
\begin{center}
\includegraphics[width=14cm]{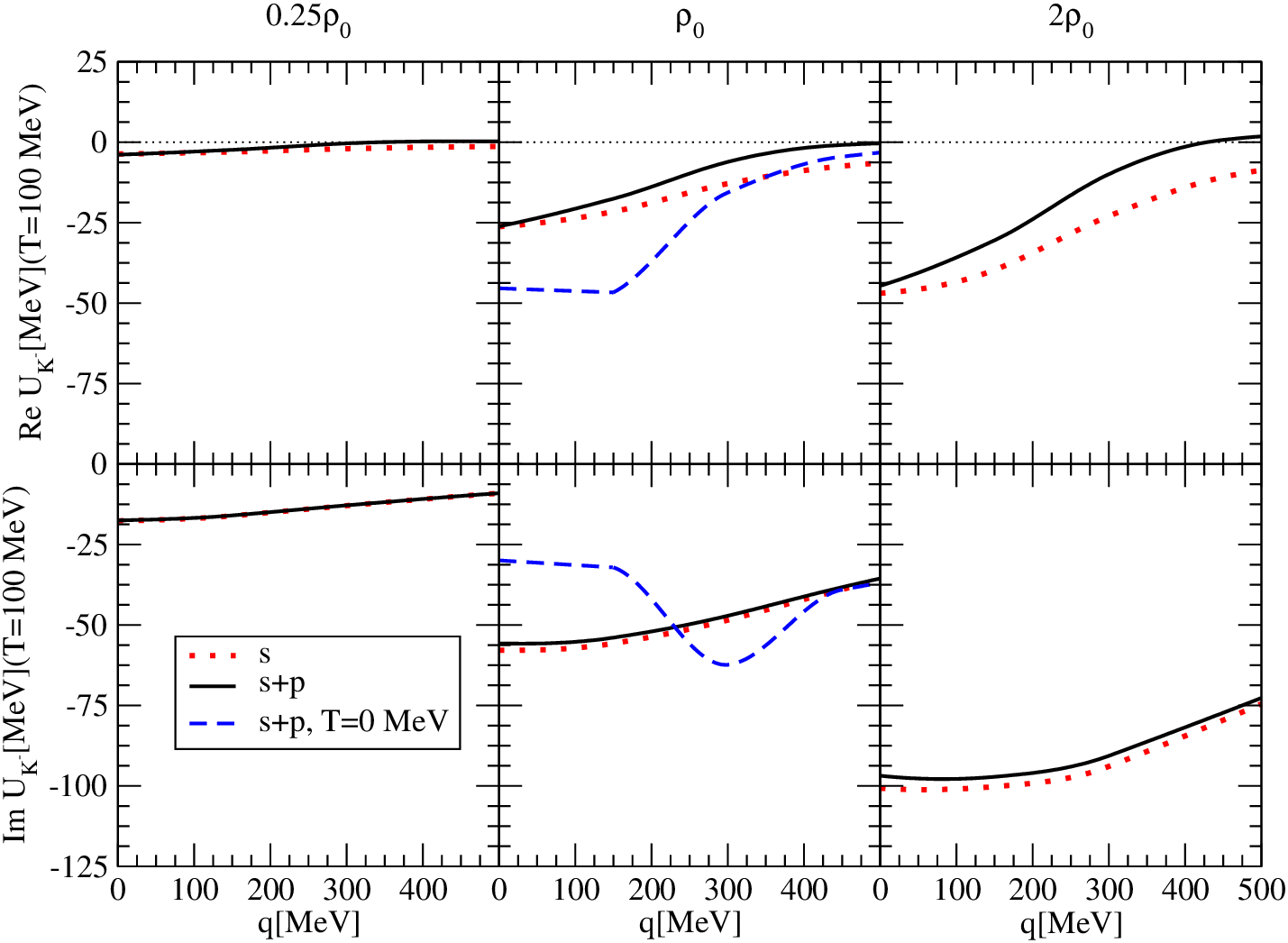}
\caption{The $\bar K$ potential for the full self-consistent calculation at
$T=100$ MeV and $0.25\rho_0$, $\rho_0$ and $2\rho_0$ as a function of momentum.
The $\bar K$ potential at $T=0$ and $\rho_0$ including $(s+p)$-waves is also
shown. }
\label{fig_upot_Kbar}
\end{center}
\end{figure}

\begin{figure}[ht]
\begin{center}
\includegraphics[width=14cm]{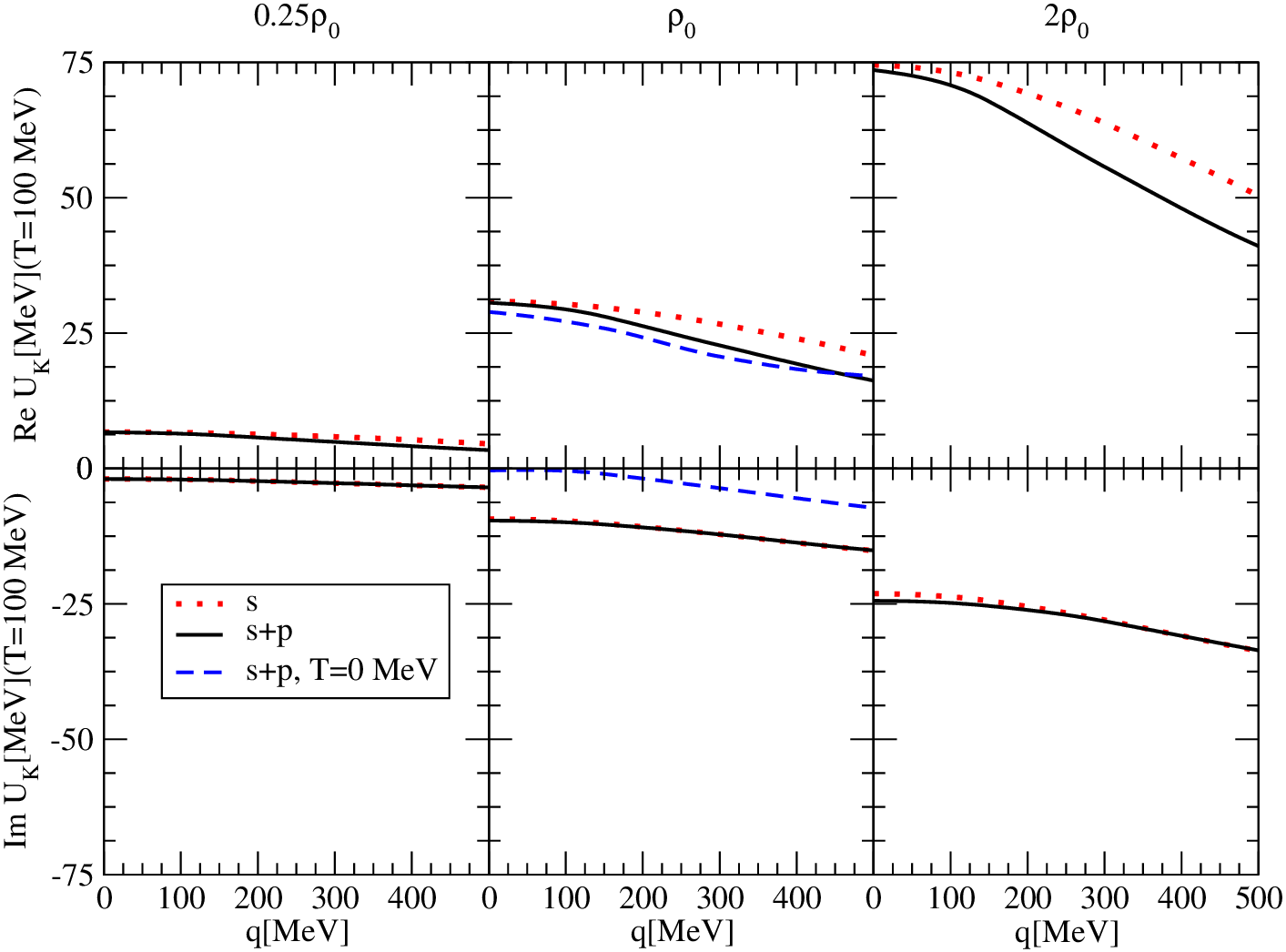}
\caption{The $K$ potential for the full self-consistent calculation at $T=100$
MeV and $0.25\rho_0$, $\rho_0$ and $2\rho_0$ as a function of momentum.  The $K$
potential at $T=0$ and $\rho_0$ including $(s+p)$-waves is also shown.}
\label{fig_upot_K}
\end{center}
\end{figure}

In this last subsection we provide the $K$ and ${\bar K}$ optical potentials at
conditions reached in heavy-ion
collisions for beam energies of the order or less than 2~AGeV, where
temperatures can reach values of $T=100$ MeV together with
densities up to a few times normal nuclear density \cite{Forster:2007qk,FOPI}.

In Figures \ref{fig_upot_Kbar} and \ref{fig_upot_K}, we show the $\bar K$ and
$K$ optical potentials at  $T=100$ MeV for different densities ($0.25\rho_0$,
$\rho_0$ and $2\rho_0$), including $s$- (dotted lines) and $(s+p)$-waves (solid
lines), as functions of the meson  momentum. In the case of $\rho_0$, we also
show the potential at  $T=0$ including $(s+p)$-waves (dashed lines). 

The real part of the $\bar K$ potential becomes more attractive as we increase
the density, going from $-4$ MeV at $0.25 \rho_0$ to $-45$ MeV at $2\rho_0$ for
$q=0$, when both $s$- and $p$-waves are included. The repulsive  $p$-wave
contributions to the $\bar K$ potential become larger as density increases,
reducing substantially the amount of attraction felt by the ${\bar K}$. Compared
to a previous self-consistent calculation using the J\"ulich meson-exchange
model \cite{TOL00,Tolos:2002ud}, here we observe a stronger dependence of the
optical potential with the ${\bar K}$ momentum.

The imaginary part of the ${\bar K}$ optical potential
at $T=100$ MeV is little affected by $p$-waves, which, as seen in
Fig.~\ref{fig_selftot}, basically modify 
the ${\bar K}$ self-energy below the quasi-particle peak. The
imaginary part of the potential shows a flat behavior at low momentum and, 
eventually, its
magnitude decreases with increasing momentum as the quasi-particle energy
moves away from the region of $YN^{-1}$ excitations.

With respect to the zero temperature case, shown for $\rho_0$ in the middle
panels of Fig.~\ref{fig_upot_Kbar}, the optical 
potential at $T=100$ MeV shows less structure. The real part amounts to 
basically half the
attraction obtained at $T=0$ MeV, while the imaginary part gets enhanced at low
momentum, due to the increase of collisional width, and reduced at high momentum, 
due to the decoupling of the ${\bar K}$ quasi-particle mode from the $\Sigma^*
N^{-1}$ one.

The $K$ meson potential changes from 7 MeV at  $0.25 \rho_0$ to 74 MeV at
$2\rho_0$ for $q=0$ MeV/c and
receives its major contribution from the $s$-wave
interaction, the $p$-wave providing a moderately attractive correction.
The imaginary part moves
from  $-2$ MeV at  $0.25 \rho_0$ to $-25$ MeV at $2\rho_0$ for $q=0$ MeV/c and
its magnitude grows moderately with increasing momentum as the available
phase space also increases.

Finite temperature affects the real part of the $K$ meson mildly, as can be seen
for $\rho_0$ from comparing the dashed and solid lines in the middle panels of
Fig.~\ref{fig_upot_K}. The magnitude of the imaginary part 
increases with temperature for all momenta, consistently with the increase of
thermically excited nucleon states.  

From our results for $\bar K$ and $K$ mesons, it is clear that $p$-wave effects
can be neglected at subnuclear densities at the level of the quasi-particle
properties. However, they become substantially important as density increases
and are also responsible for a considerable amount of the strength at low energies in the spectral
function.

\section{Summary, conclusions and outlook}
\label{sec:Conclusion}

We have obtained the $\bar K$ and $K$ self-energies in symmetric nuclear matter
at finite temperature from a chiral unitary approach, which incorporates the
$s$- and $p$-waves of the kaon-nucleon interaction. At tree level, the $s$-wave
amplitude is obtained from the Weinberg-Tomozawa term of the chiral Lagrangian.
Unitarization in coupled channels is imposed by solving the Bethe-Salpeter
equation with on-shell amplitudes. The model generates dynamically the  $\Lambda
(1405)$ resonance in the $I=0$ channel. The in-medium solution of  the $s$-wave
amplitude, which proceeds by a re-evaluation of the meson-baryon loop function,
accounts for Pauli-blocking effects, mean-field binding on the nucleons and
hyperons via a temperature-dependent $\sigma-\omega$ model, and the dressing of
the pion and kaon through their corresponding self-energies. This requires a
self-consistent evaluation of the $K$ and $\bar K$ self-energies.  The $p$-wave
self-energy is accounted for through the corresponding  hyperon-hole ($YN^{-1}$)
excitations. Finite temperature expressions have been obtained in the Imaginary
Time Formalism, giving a formal justification of some approximations typically
done in the literature and, in some cases, improving upon the results of
previous works. For instance, in this formalism, the Lindhard function of
$YN^{-1}$ excitations automatically accounts for Pauli blocking on the excited
hyperons and satisfies the analytical constraints of a retarded self-energy.

The $\bar K$ self-energy and, hence, its spectral function show a strong mixing
between the quasi-particle peak and the $\Lambda(1405)N^{-1}$ and  $YN^{-1}$
excitations. The effect of the $p$-wave $YN^{-1}$ subthreshold excitations is
repulsive for the $\bar K$ potential, compensating in part the attraction
provided by the $s$-wave ${\bar K} N$ interaction.  Temperature softens the
$p$-wave changes on the spectral function at the quasi-particle energy. On the
other hand, together with the $s$-wave mechanisms, the $p$-wave self-energy
provides a low-energy tail which spreads the spectral function considerably, due
to the smearing of the Fermi surface for nucleons. Similarly, the size of the
imaginary part of the potential decreases with momentum,  as the $\bar K$ mode
decouples from subthreshold absorption mechanisms.

The narrow $K$ spectral function dilutes with density and temperature as the
number of collisional $KN$ states is increased. A moderate repulsion, coming
from the dominant $s$-wave self-energy, moves the quasi-particle peak to higher
energies in the hot and dense medium. The absence of resonant states close to 
threshold validates the use of the low-density theorem for the $K$ optical 
potential approximately up to saturation density. The inclusion of $p$-waves has a mild attractive effect on
the $K$ self-energy and potential, which results from $YN^{-1}$ excitations in
crossed kinematics.

The properties of strange mesons at finite temperature for densities of 2-3
times normal nuclear matter have been object of intensive research in the
context of relativistic heavy-ion collisions at beam energies below 2~AGeV
\cite{Forster:2007qk}. The comparison between the experimental results on
production cross sections, energy distributions and polar angle distributions,
and the different transport-model calculations has lead to several important
conclusions, such as the coupling between the $K^-$ and the $K^+$ yields by
strangeness exchange and the fact that the $K^+$ and $K^-$ mesons exhibit
different freeze-out conditions. However, there is still debate on the influence
of the kaon-nucleus potential on those observables. The in-medium modifications
of the $\bar K$ and $K$ properties devised in this paper could be used in
transport calculations and tested against the data from the current experimental
programs in heavy ions \cite{Forster:2007qk,FOPI}.

The fact that the $\bar{K}$ spectral function spreads to low 
energies, even at finite momentum, may have relevant implications 
on the phenomenology of the $\phi$
meson propagation and decay in a nuclear medium.
The reduced phase space for the
dominant decay channel in vacuum, $\phi \to \bar K K$, makes the $\phi$ meson
decay width a sensitive probe of kaon properties in a hot and dense medium (the
$p$-wave nature of the $\phi \bar K K$ coupling further enhances this
sensitivity). In
\cite{Oset:2000eg,Cabrera:2002hc} the $\phi$ meson mass and decay width in
nuclear matter were studied from a calculation of the $\bar K$ and $K$
self-energies in a chiral unitary framework similar to the present work 
(the most
relevant differences and novelties introduced in this work have been discussed
in previous sections). The overall attraction of the $\bar K$ meson together
with a sizable broadening of its spectral function (which reflects the fate of
the $\Lambda (1405)$ in a nuclear medium), induced a remarkable increase of
$\Gamma_{\phi}$ of almost one order of magnitude at $\rho=\rho_0$ as compared to
the width in free space, as several decay mechanisms open in the medium such as
$\phi N \to K Y$ and $\phi N \to K \pi Y$. The LEPS Collaboration
\cite{Ishikawa:2004id} has
confirmed that the $\phi$ meson width undergoes strong modifications in the
medium from the study of the inclusive nuclear $\phi$ photoproduction reaction
on different nuclei. The observed effects even surpass the sizable modifications
obtained in \cite{Oset:2000eg,Cabrera:2002hc} and predicted for the $\phi$
photoproduction reaction in \cite{Cabrera:2003wb}.

At finite temperature (CERN-SPS, SIS/GSI and FAIR/GSI
conditions), despite the $\bar K$ peak return towards its free position, we
expect a similar or even stronger broadening of the $\phi$ meson, as $S_{\bar
K}$ further dilutes in the medium effectively increasing the available phase
space. Note, in addition, that the presence of thermally excited mesons induces
''stimulated'' $\phi \to \bar K K$ decays (as well as diffusion processes)
\cite{Gale:1990pn,Haglin:1994ap}.
Since Bose enhancement is more effective on the lighter modes of
the system, the low energy tail of the $\bar K$ spectral
function may contribute substantially to the $\phi$ decay width.

At RHIC and LHC conditions, the hot medium is expected to have lower net
baryonic composition. One may conclude, as a consequence, that the contribution
from interactions with baryons will be smaller. 
The relevance of baryonic
density effects even at high temperatures has been stressed for the $\rho$ and
$\phi$ meson clouds in hot and dense matter
\cite{Smith:1997xu,Rapp:1999ej,Rapp:2000pe}.
In the case of $\phi \to \bar K K$ decays, a finite density of
antibaryons allows the $K$ to interact with the medium through the
charge-conjugated mechanisms described here for the $\bar K$, and viceversa
(in the limit of $\mu_B =0$ the $\bar K$ and $K$ self-energies are identical).
At small net baryon density, whereas the effective contribution from the real
parts of the $K$, $\bar K$ self-energies tend to vanish, it is not the case for
the imaginary parts, which are always cumulative.
Thus even at small baryonic chemical potential, the presence of
antibaryons makes up for the loss of reactivity from having smaller nuclear
densities.
Additionally, the
relevance of kaon interactions with the mesonic gas (ignored in this work)
becomes manifest in this regime, as it has been pointed out in
\cite{AlvarezRuso:2002ib,Holt:2004tp,Faessler:2002qb,Santini:2006cm}.
This mechanisms, together with Bose
enhancement of $\bar K K$ decays point towards a sizable increase of the $\phi$
reactivity even at very high temperatures.

Therefore, we plan to study the influence of the $\bar K K$ cloud on the
properties of the $\phi$ meson in a nuclear medium at finite temperature
\cite{Dani}, extending our previous analysis for cold nuclear matter
\cite{Oset:2000eg,Cabrera:2002hc,Cabrera:2003wb}.
Such changes on the $\phi$ meson properties are a matter of interest in the
current and future experimental heavy-ion programs \cite{FOPI,HADES,CBM}. In
particular, the future FAIR facility at GSI will devote special attention to the
in-medium vector meson spectral functions. The HADES experiment will operate at
higher beam energy of the order of 8-10 AGeV, providing complementary
information on the spectral function evolution of vector mesons to the current
research program. On the other hand, CBM will measure the in-medium spectral
functions of short lived vector mesons directly by their decay into dilepton
pairs.

With this work we expect to pave the understanding of kaon properties in hot and
dense matter and provide an essential ingredient for the $\phi$-meson
phenomenology in heavy-ion collisions. Our results are based on a self-consistent
many-body calculation at finite temperature which relies on a realistic model of the kaon nucleon
interaction thoroughly confronted to the kaon nuclear phenomenology.

\section{Acknowledgments}

We thank E. Oset for useful discussions. We also thank R.~Rapp for helpful
discussions and comments at the initial stage of the project. This work is
partly supported by the EU contract FLAVIAnet MRTN-CT-2006-035482, by the
contract FIS2005-03142 from MEC (Spain) and FEDER and by the Generalitat de
Catalunya contract 2005SGR-00343. This research is part of the EU Integrated
Infrastructure Initiative Hadron Physics Project under contract number
RII3-CT-2004-506078. L.T. wishes to acknowledge support from the BMBF project
``Hadronisierung des QGP und dynamik von hadronen mit charm quarks''  (ANBest-P
and BNBest-BMBF 98/NKBF98). D.C. acknowledges support from the "Juan de la
Cierva" Programme (Ministerio de Educaci\'on y Ciencia, Spain).

\appendix

\section{Finite-temperature Lindhard functions}
\label{app-Linds}
We quote here some Lindhard function expressions in the ITF to
point out the main differences with previous evaluations.

\subsection{$YN^{-1}$ excitations}

In the ITF, the $YN^{-1}$ Lindhard function reads

\begin{eqnarray}
\label{eq:Lind-rel-YN} {\mathcal U}_{Y N^{-1}}(\omega_n,\vec{q};
T) = 2\, \int \frac{d^3p}{(2\pi)^3}
\frac{n_N(\vec{p},T)-n_Y(\vec{p}+\vec{q},T)} {{\rm i}\omega_n +
E_N(\vec{p},T) - E_Y(\vec{p}+\vec{q},T)} \ ,
\end{eqnarray}
where $\omega_n$ is a bosonic Matsubara frequence
(${\rm i} \omega_n={\rm i}2n\pi T$) and the factor 2 stands for
spin degeneracy.
Consistently with the approximations employed in this work, we
have only kept the positive energy part of the baryon propagators,
while keeping the baryon energies fully relativistic and
containing also mean-field binding potentials. The nucleon and
hyperon Fermi distributions, $n_{N,Y}(\vec{p},T)=
[e^{(E_{N,Y}(\vec{p},T)-\mu_B)/T}+1]^{-1}$, depend on the
temperature and baryon chemical potential, so that for fixed $T$
and $\mu_B$ the nucleon and hyperon densities are given by
\begin{equation}
\label{eq:NandYdensities} \rho_N = \nu_N\, \int
\frac{d^3p}{(2\pi)^3} n_N(\vec{p},T) \,\,\, , \ \ \ \rho_Y = \nu_Y
\, \int \frac{d^3p}{(2\pi)^3} n_Y(\vec{p},T) \,\,\, ,
\end{equation}
with $\nu_B$ the corresponding spin-isospin degeneracy factors,
namely $\nu_N=4$ and $\nu_Y = (2,6,12)$ for $Y=(\Lambda, \Sigma,
\Sigma^*)$. All the hyperons are considered as stable particles.

Note that, differently from the zero temperature case shown below, at finite
temperature there are occupied hyperon states and therefore the
$Yh$ excitation is suppressed by hyperon Pauli blocking, as it is
evident from the hyperon distribution, $n_Y$, which appears
subtracting in the numerator of the Lindhard function. This can be
seen explicitly by rewriting $n_N-n_Y$ as  $n_N (1+n_Y) - n_Y
(1+n_N)$.

Analytical continuation to real energies (${\rm i} \omega_n \to
q_0 + {\rm i} \varepsilon$) gives the following expressions for
the real and imaginary parts of the finite temperature $YN^{-1}$
Lindhard function (including mean-field binding potentials),
\begin{eqnarray}
\label{eq:Lind-rel-YN-elaborate} {\rm Re}\, U_{Y
N^{-1}}(q_0,\vec{q}; T) &=& \frac{1}{2\pi^2} \int_0^{\infty} dp\,
p^2\,\,{\cal P} \int_{-1}^{+1} du \,
\frac{n_N(\vec{q},T)-n_Y(\vec{p}+\vec{q},T)} {q_0 + E_N(\vec{p},T)
- E_Y(\vec{p}+\vec{q},T)} \,\,\, ,
\nonumber \\
{\rm Im}\, U_{Y N^{-1}}(q_0,\vec{q}; T) &=& - \pi \,
\frac{1}{2\pi^2} \int_0^{\infty} dp\, p^2\,
\frac{q_0+E_N(\vec{p},T) - \Sigma_Y}{p\,q} \,
\nonumber \\
&\times& [n_N(\vec{q},T)-n_Y(\vec{p}+\vec{q},T)]_{u_0} \, \theta
(1-|u_0|) \, \theta (q_0 +E_N(\vec{p},T) - \Sigma_Y) \,\,\,,
\nonumber \\
\end{eqnarray}
with $u_0 \equiv u_0 (q_0,q,p) = [(q_0+E_N-\Sigma_Y)^2 - (M_Y^*)^2
- p^2 - q^2]/(2\,p\,q)$, where here $q$, $p$ refer to the modulus
of the corresponding three-momentum.

In the limit of zero temperature and fixed baryon chemical
potential, $\mu_B$, which then coincides with the nucleon Fermi energy,
$E_F=E_N(p_F)$, with $p_F$ the Fermi momentum, we have
\begin{equation}
\label{T0limit} n_N(\vec{p},T) \to n_N(\vec{p}\,) = \theta (p_F-p) \
, \ n_Y(\vec{p},T) \to 0 \ .
\end{equation}
The $YN^{-1}$ Lindhard function then reads
\begin{equation}
\label{LindYN-T0} U_{Y N^{-1}}(q_0,\vec{q}; \rho) = 2\, \int
\frac{d^3p}{(2\pi)^3} \frac{n_N(\vec{p}\,)} {q_0 + E_N(\vec{p}\,)
- E_Y(\vec{p}+\vec{q}\,) + {\rm i} \varepsilon} \ ,
\end{equation}
and $\rho = 2\, p_F^3 / 3\,\pi^2$. In
\cite{Oset:2000eg,Cabrera:2002hc,Tolos:2002ud} analytical
expressions are provided for the non-relativistic Fermi gas (i.e.,
$E_B(\vec{p}\,)=M_B+\vec{p}\,^2/2M_B$ above), which we quote here
for completeness,
\begin{eqnarray}
\label{LindYN-T0-norel}
{\rm Re}\, U^{{\rm nr}}_{YN^{-1}}(q_0,\vec{q};\rho)
&=& \frac{3}{2} \rho
\frac{M_Y}{q p_F} \Bigg\{  z+ \frac{1}{2}(1-z^2) {\rm ln}
\frac{|z+1|}{|z-1|} \Bigg\}  \ ,
\nonumber \\
{\rm Im}\, U^{{\rm nr}}_{YN^{-1}}(q_0,\vec{q};\rho)
&=&-\frac{3}{4} \pi \rho
\frac{M_Y}{q p_F}  \lbrace (1-z^2) \theta(1-|z|) \rbrace \ ,
\end{eqnarray}
with
\begin{equation}
z=\frac{M_Y}{q p_F}\left\{q_0-\frac{\vec{q}\,^2} {2
M_Y}-(M_Y-M_N)\right\} \ .
\end{equation}
We note that terms proportional to ($M_Y^{-1}-M_B^{-1}$) in the denominator of
Eq.~(\ref{LindYN-T0}) are neglected when doing the angular
integration to arrive to Eq.~(\ref{LindYN-T0-norel}).

In the literature, one often finds approximate expressions for the
finite temperature Lindhard function that have been obtained from
extensions of the former $T=0$ equations. In \cite{Tolos:2002ud}
the finite temperature generalization of Eq.~(\ref{LindYN-T0}) was
obtained by replacing the nucleon occupation number, $n_N(\vec{p}\,)
= \theta (p_F-p)$,  by the corresponding Fermi distribution at
finite temperature, $n_N(\vec{p}\,) \to n_N(\vec{p},T)$. Analytical
expressions (up to a momentum integration) read
\begin{eqnarray}
\label{LindYN-Tfinita-norel}
{\rm Re}\, U^{{\rm nr}}_{YN^{-1}}(q_0,\vec{q};T)
&=& \frac{1}{\pi^2}
\frac{M_Y}{q} \int dp \ p \ n_N(\vec{p},T)
\ {\rm ln} \frac{|z+1|}{|z-1|}  \ ,
\nonumber \\
{\rm Im}\, U^{{\rm nr}}_{YN^{-1}}(q_0,\vec{q};T)
&=& - \frac{1}{\pi}  \, T
\frac{M_N \, M_Y}{q} \ {\rm ln} \frac{1}{1-n_N(\vec{p}_m,T)}\ ,
\end{eqnarray}
with
\begin{eqnarray}
z&=&\frac{M_Y}{q p}\left\{q_0-\frac{\vec{q}\,^2}
{2 M_Y}-(M_Y-M_N)\right\} \ , \nonumber \\
p_m&=&\frac{M_Y}{q} \left | q_0-(M_Y-M_N)-\frac{\vec{q}\,^2}{2 M_Y}
\right | \ .
\end{eqnarray}

\subsection{$ph$ and $\Delta h$ excitations}

The evaluation of the $ph \equiv NN^{-1}$ Lindhard function at
finite temperature and density can be found, for instance, in
\cite{Rapp-rho-finiteT,mattuck}. Analytical continuation to real
energies from the ITF expression leads to
\begin{eqnarray}
\label{eq:Lind-rel-ph}
U_{N N^{-1}}(q_0,\vec{q}; T) =
\nu_N \, \int \frac{d^3p}{(2\pi)^3} \frac{n_N(\vec{p},T)-n_N(\vec{p}+\vec{q},T)}
{q_0 + {\rm i}\varepsilon + E_N(\vec{p},T) - E_N(\vec{p}+\vec{q},T)}
\,\,\, ,
\end{eqnarray}
with $\nu_N=4$.
Similarly, for the $\Delta h$ Lindhard function we arrive at
\begin{eqnarray}
\label{eq:Lind-rel-Dh}
U_{\Delta N^{-1}}(q_0,\vec{q}; T)
&=&
\nu_{\Delta} \,
\int \frac{d^3p}{(2\pi)^3}
\left[
\frac{n_N(\vec{p},T)-n_{\Delta}(\vec{p}+\vec{q},T)}
{q_0 + {\rm i} \frac{\Gamma_{\Delta}(q_0,\vec{q}\,)}{2}
+ E_N(\vec{p},T) - E_{\Delta}(\vec{p}+\vec{q},T)}
\right.
\nonumber \\
&+&
\left.
\frac{n_{\Delta}(\vec{p},T)-n_N(\vec{p}+\vec{q},T)}
{q_0 + {\rm i} \frac{\Gamma_{\Delta}(-q_0,\vec{q}\,)}{2}
+ E_{\Delta}(\vec{p},T) - E_N(\vec{p}+\vec{q},T)}
\right]
\,\,\, ,
\end{eqnarray}
which we have written explicitly in terms of direct ($\Delta
N^{-1}$) plus crossed ($N\Delta^{-1}$) contributions. For
convenience, the $\pi N\Delta$ coupling is absorbed in the
definition of $U_{\Delta N^{-1}}$ and thus $\nu_{\Delta} =
\frac{16}{9} (f_{\Delta} / f_N)^2$. Note that in
Eq.~(\ref{eq:Lind-rel-Dh}) we have accounted for the decay width
of the $\Delta$ resonance. A realistic treatment of the $\Delta h$
mechanism should account for the full energy dependent $\Delta$
decay width into its dominant channel, $\pi N$. We implement the
$\Delta$ decay width using the formulae in \cite{Oset:1989ey}, in
which $\Gamma_{\Delta}$ only depends on the pion energy and
momentum (a more detailed study of the in-medium $\Delta$
self-energy at finite temperature and density has been reported in
\cite{vanHees:2004vt}). $\Gamma_{\Delta}(q_0,\vec{q}\,)$ accounts
for both direct and crossed kinematics and hence the retarded
property of $U_{\Delta N^{-1}}$ is preserved. We have not
accounted for binding effects on the nucleon and $\Delta$ in the
pion self-energy.  First, in the $ph$ excitation the baryonic
potentials cancel to a large extent. Second, the binding
potentials for the $\Delta$ resonance are not well known
experimentally and we omit them. Therefore, for consistence, we do
not dress the nucleon in the $\Delta h$ excitation either.

The $T\to 0$ limit (at nuclear matter conditions) of $U_{NN^{-1}}$ and
$U_{\Delta N^{-1}}$ can be easily obtained with similar prescriptions as in the
$YN^{-1}$ case, namely, $n_N(\vec{p},T)\to n_N(\vec{p}\,)$ and
 $n_{\Delta} \to 0$.
Analytic expressions for the non-relativistic $ph$ and $\Delta h$ Lindhard
functions can be found in \cite{Oset:1989ey}.

\section{$s$-wave self-energy from $T_{{\bar K}(K) N}$}
\label{app-swave}

We derive in this section a general expression for the contribution to the  kaon
self-energy from the effective in-medium $\bar K (K) N$ scattering amplitude at
finite temperature. Let us denote by $T_{{\bar K}(K)N}$ the isospin averaged
kaon nucleon scattering amplitude. The kaon self-energy, $\Pi_{{\bar K}(K)N}$,
follows from closing the nucleon external lines and, following the Feynman rules
in the ITF, reads
\begin{equation}
\label{s-wave-ITF}
\Pi_{{\bar K}(K)N} (\omega_n,\vec{q}; T) = T\, \sum_{m=-\infty}^{\infty}
\int \frac{d^3p}{(2\pi)^3} \, \frac{1}{{\rm i} W_m - E_N (\vec{p}\,)}
\, T_{{\bar K}(K)N} (\omega_n + W_m , \vec{P} ; T)
\ ,
\end{equation}
where $\omega_n$ and $W_m$ are bosonic and fermionic Matsubara frequencies,
respectively, ${\rm i} \omega_n = {\rm i} 2n\pi T$ and
${\rm i} W_m = {\rm i} (2m+1)\pi T + \mu_B$.
The sum over the
index $m$ is not straightforward since $T_{{\bar K}(K)N}$ depends on $m$ in a
non-trivial way. To skip this complication, we can invoke a spectral
representation for the $T$-matrix (inherited from the analytical structure of
the meson-baryon loop function) and we have
\begin{eqnarray}
\label{s-wave-ITF-2}
\Pi_{{\bar K}(K)N} (\omega_n,\vec{q}; T)
&=&
-T\, \sum_{m=-\infty}^{\infty}
\int \frac{d^3p}{(2\pi)^3} \,
\frac{1}{\pi}\int_{-\infty}^{\infty} d\Omega \,
\frac{{\rm Im}\,T_{{\bar K}(K)N}(\Omega,\vec{P};T)}
{[{\rm i} W_m - E_N (\vec{p}\,)] [{\rm i} \omega_n + {\rm i} W_m - \Omega]}
\nonumber \\
&=&
- \int \frac{d^3p}{(2\pi)^3} \,
\frac{1}{\pi}\int_{-\infty}^{\infty} d\Omega \,
\frac{{\rm Im}\,T_{{\bar K}(K)N}(\Omega,\vec{P};T)}
{{\rm i} \omega_n - \Omega + E_N (\vec{p}\,)}
\, [n_N(\vec{p},T) - n(\Omega,T)] \, ,
\nonumber \\
\end{eqnarray}
with $n(\Omega,T) = [e^{(\Omega-\mu_B)/T}+1]^{-1}$ here.
The former result, after
continuation into the real energy axis (${\rm i}\omega_n \to q_0+{\rm
i}\varepsilon$), provides the thermal kaon self-energy evaluated from the kaon
nucleon scattering amplitude. Note that it includes a Pauli blocking correction
term, $n(\Omega,T)$, convoluted with the spectral strength from the imaginary
part of the $T$-matrix. At the region in which the principal value of the
spectral integration gets its major contribution, $\Omega \approx q_0 +
E_N(\vec{p}\,)$, the fermion distribution $n(\Omega,T)$ behaves as a slowly
varying exponential tail (for the present temperatures under study). We can
approximate this term by a constant, namely, $n(\Omega,T)\simeq n(q_0 +
E_N(\vec{p}\,),T)$ and take it out of the integral. The dispersion
integral over $\Omega$ then recovers the full amplitude $T_{{\bar K}(K)N}$ and
the self-energy can be approximated by:
\begin{equation}
\label{s-wave-ITF-3}
\Pi_{{\bar K}(K)N} (q_0+{\rm i}\varepsilon,\vec{q}; T) =
\int \frac{d^3p}{(2\pi)^3} \,
T_{{\bar K}(K)N} (q_0 + E_N(\vec{p}\,),\vec{P};T) \,
[n_N(\vec{p},T) - n(q_0 + E_N(\vec{p}\,),T)]
\ .
\end{equation}
Note that this procedure is exact for the imaginary part.
Eq.~(\ref{eq:selfd}) follows from the former result by neglecting the Pauli
blocking correction on the fermion degrees of freedom excited in the kaon
nucleon amplitude (note that in the isospin zero amplitude the strength peaks
around the $\Lambda(1405)$ resonance, and thus one expects this correction to be
small with respect to that on the nucleon).

\end{document}